\documentclass[11pt]{article}
\usepackage{amsmath,amscd,amssymb,graphicx}
\newcommand{\wwr}{|\widehat{0}\rangle}
\newcommand{\xxr}{|\widehat{1}\rangle}
\newcommand{\wwl}{\langle\widehat{0}}
\newcommand{\xxl}{\langle\widehat{1}}
\newcommand{\Nhat}{\widehat{\cal N}}

\newcommand{\be}{\begin{equation}}
\newcommand{\ee}{\end{equation}}
\newcommand{\bea}{\begin{eqnarray}}
\newcommand{\eea}{\end{eqnarray}}

\begin{document}

\title{Topological order from quantum loops and nets}

\author{Paul Fendley\\
\\
All Souls College and the Rudolf Peierls Centre for Theoretical
Physics,\\ University of Oxford, 1 Keble Road,  OX1 3NP, UK;\\
and Department of Physics, University of Virginia,
Charlottesville, VA 22904-4714 USA
}
 
\date{\today} 

\maketitle

\begin{abstract} 
I define models of quantum loops and nets that have ground states with
topological order. These make possible excited states comprised of
deconfined anyons with non-abelian braiding.  With the appropriate
inner product, these quantum loop models are equivalent to net models
whose topological weight involves the chromatic polynomial. A simple
Hamiltonian preserving the topological order is found by exploiting
quantum self-duality. For the square lattice, this Hamiltonian has
only four-spin interactions.

\end{abstract}

\section{Introduction}

Condensed-matter systems with anyonic excitations have been the
subject of intense study recently, one reason being their potential
application to topological quantum computation \cite{DFNSS}. Anyonic
excitations can occur in many-body systems with {\em
  fractionalization}, i.e.\ where the quasiparticles of the system
have quantum numbers which are fractions of the underlying
``fundamental'' excitations. For example, in the fractional quantum
Hall effect, quasiparticles of charge $e/3$ and $e/5$ have been
observed, even though the system is comprised purely of electrons.
The Hall effect has given the only unambiguous experimental
realization of fractionalized excitations, but since this behavior is
so novel, and the consequences potentially so dramatic, it is
important to find different sorts of models exhibiting this behavior.
Thus considerable theoretical attention also has been devoted to
finding magnetic models exhibiting anyonic excitations.

It is now
well understood how abelian fractionalized excitations can occur in
relatively simple spin systems, e.g.\ the ``toric code''
\cite{Kitaev97}, and quantum dimer models \cite{MS}. 
For topological quantum computation, however, it is essential that the
excitations not only be fractionalized, but also have non-abelian
braiding.  Magnetic models with this behavior are much trickier to
find. For a model containing non-abelian anyons to be mathematically
consistent, the braiding and fusing properties of the anyons must
satisfy very intricate constraints \cite{DFNSS,Preskill}. At a glance, one
might think it impossible for the quasiparticles in any
physical system to satisfy these properties. Nevertheless, we know
from the fractional quantum Hall effect that these properties do
indeed arise as a
consequence of well-motivated (and well-checked numerically) model
wavefunctions \cite{MR}. It is thus reasonable to hope that non-abelian
excitations can arise in a physically-relevant system.

At the moment, however, magnetic models with non-abelian anyons do not
arise as naturally as those coming from the Hall effect or other
chiral systems. Thus attempts to find them have mainly been reverse
engineering \cite{Freedman01}. Namely, first one finds a
mathematically-consistent set of braiding and fusion relations; in
mathematics this is called a modular tensor category, while in physics
these rules are familiar from two-dimensional conformal field
theory. Then one attempts to fine-tune a Hamiltonian to obtain this
braiding and fusing. With sufficiently complicated interactions, it is
indeed possible to build this structure into a Hamiltonian, for
example in ``string-net'' models \cite{LevinWen}. Moreover, in at
least one special case, it is also possible to find a simple (although
highly fine-tuned) Hamiltonian whose excitations involve Majorana
fermions, and when time-reversal symmetry is broken, non-abelian
braiding \cite{Kitaev2}. These Hamiltonians therefore provide valuable
proofs of principle.

Fractionalized excitations (abelian and non-abelian) in magnetic
models generally arise because they have non-trivial topological
characteristics. Braiding is an inherently topological property --
the statistics depends only on the topology of the paths in
configuration space for taking the particles around each other.  In
order to find more realistic models, it is therefore desirable to find
ones which incorporate directly the {\em topological} properties of
non-abelian braiding, not only the algebraic ones. To this end,
substantial attention has been given to {\em quantum loop models}
\cite{Freedman01}. These have the nice feature that the necessary
topological properties of non-abelian braiding are built in from the
beginning.

More precisely, an essential property for having deconfined anyons is
that the ground state can be expressed as a superposition of
states, each of which is characterized by a set of
geometric objects. In
quantum loop models each configuration in the ground-state sum is
described by a set of closed loops on the links of some lattice.  A
more general geometric object central to this paper is the {\em net},
where ``branching'' (e.g.\ trivalent vertices) is allowed, but the
geometric objects still do not have ends.  Precisely, the ground state
$|\Psi\rangle$ can be written as a sum over geometric
objects ${\cal G}$
\begin{equation}
|\Psi\rangle = \sum_{\cal G} w({\cal G}) |{\cal G}\rangle\ .
\label{PsiG}
\end{equation}
In the quantum loop models discussed below, the weight
depends on the number of loops, while for the nets, it will depend on
the chromatic polynomial of a (dual) graph.

A tell-tale sign of a ground state described as a sum over geometric
objects is that the number of ground states
depends on the topology of two-dimensional space. Such behavior is
often referred to as ``topological order'', and such a system is said
to be in a ``topological phase''. For example, in the toric code and
in the triangular-lattice quantum dimer model \cite{RK}, there is only one
ground state when space is topologically a sphere, but there are four
when space is topologically a torus. The latter is best understood in
terms of loops: the four cases correspond to having an even or odd
number of loops wrap around each of the two cycles of the torus.

When the ground state has this structure, it is then easy to
understand how fractionalized excitations arise. 
A state with a pair of anyons looks like the ground state (e.g.\ a sum
over configurations with closed loops), except at the locations of the
anyons. Each anyon corresponds to the end of a loop or net, so that
each pair of anyons is attached by a segment of loop or net. In more
picturesque terms, ``cutting a loop'' results in an anyon pair. Since
away from the anyons themselves, the state looks like the ground
state, the energy density of such a state is localized around the
locations of the anyons. Such a configuration is exactly what one
would describe as a two-quasiparticle state. The novelty here is that
in each configuration in this sum, the quasiparticles are attached by
segments of loop, no matter how far apart they are. This is why it is
possible for non-trivial braiding to occur: when quasiparticles are
brought around each other, there are strings attached! The segments of
loop attached must go through each other when they are attached to
particles being braided. When there are multiple states with the same
energy and quantum numbers, the braiding is then described by a
matrix, opening up the possibility of non-abelian braiding.

This picture, introduced in \cite{Kitaev97,MS,Freedman01} and developed
further in papers like \cite{FNS,FF,Fidkowski}, is very
compelling. Moreover, it is not hard to find model Hamiltonians which
have the properties described above. 

The difficulties in implementing this proposal revolve around ensuring
that the anyons are {\em deconfined}: their energy should not increase
as they are moved apart from each other. In terms of the geometric
picture, this means that effectively the loop or net segments should
have effectively no energy per length. This seems like it should be
automatic, given that the Hamiltonian is chosen so that the ground
state is a sum over loop configurations of all lengths, and the
weights only depend on topological properties of the loops. However,
it is well understood in two-dimensional classical loop models how a
confinement scale can be introduced. A classical loop model must be
critical in order for loops of all length scales to contribute to the
partition function. Nevertheless, when the weight per loop exceeds a
certain value, it is impossible for there to be a critical point with
deconfined loops \cite{Nienhuis}. Below I refer to the analogous
problem in quantum loop models as the $d=\sqrt{2}$ barrier.

In this paper I explain how to go beyond the $d=\sqrt{2}$ barrier by
specifying the inner product on the Hilbert space appropriately.  This
inner product is local and has desirable topological properties.  A
striking consequence is that the loop models are
equivalent to the net models of \cite{FF}. Namely, the appropriate
inner product for quantum loop models leads naturally to describing
the degrees of freedom in terms of Ising spins. Moreover, I will show
how in the ground state these Ising spins form nets whose weights
depend on a topological quantity, the chromatic polynomial of the dual
of the net \cite{FF}.  The string-net models of
\cite{LevinWen,Fidkowski} are generalizations of these net models; the
string-net models have the advantage of being exactly solvable, while
the models here have a much simpler Hamiltonian. The exact solution of
the string-net models means that one can prove anyons are deconfined
there; I will argue here that the same topological phase occurs in
the models here.

As discussed above, a prime desire in quantum models with geometric
degrees of freedom is to find
topological order and fractionalized excitations. Since 
these arise only from strong correlations, this behavior
cannot be understood by using traditional perturbative
techniques. Thus the main method of making the problem tractable has
been to utilize model Hamiltonians, in the hope that these capture the
essential physics of the entire phase.

A key tool in finding a Hamiltonian comes from exploiting the {\em
quantum self-duality} of the net models introduced here.  The ground
state can be described in terms of geometric degrees of freedom in two
separate but equivalent ways, on a lattice and its dual.  This is a
consequence of the fact that the anyon fusion matrix describes the
change of basis to the dual.  The quantum self-duality makes it easy
to find a Hamiltonian with $|\Psi\rangle$ as its ground state.  This
Hamiltonian involves only interactions around a vertex and a face, so
that for the square lattice this requires only four-spin interactions
(thinking of the net variable on each link as a ``spin''.) This
Hamiltonian is therefore considerably simpler than that of other known
time-reversal-invariant models with non-abelian statistics in the
spectrum.  (The string-net models \cite{LevinWen}, for example,
require 12-spin interactions.)

In section \ref{sec:naive}, I define the quantum loop models
considered, and explain precisely what the difficulties with using the
naive inner product on the Hilbert space.  The modified inner product
is described in section \ref{sec:cracking}, where I explain how the
classical loop model results of \cite{FJ} allow the $d=\sqrt{2}$
barrier to be cracked in quantum loop models.  In section
\ref{sec:nets}, the map to the net models is given in detail.  Several
examples are given in section \ref{sec:examples}, making contact both
with the toric code and with string-net models. The concept of quantum
self-duality is described in section \ref{sec:selfdual}, and used
to find Hamiltonians with the desired ground state. How quantum-group
algebras enter into this picture is explained in section \ref{sec:qga}. Some
generalizations are briefly discussed in section \ref{sec:generalizations}.
The explicit derivation of the topological weight of the net in terms
of the chromatic polynomial is presented in an appendix.

Some of the results of sections 2 and 3 were discussed in
arXiv:0711.0014. This paper renders that preprint obsolete, and changes the
notation considerably.

\section{Quantum loop models and their inner product(s)}
\label{sec:naive}

As discussed in the introduction, the desired ground
state in a topological phase is a sum over geometric objects. The
simplest example of such is to make the geometric objects self- and
mutually-avoiding {\em loops} \cite{Freedman01}. 
The Hilbert space of a quantum loop model is spanned by loop
configurations on some lattice; each loop configuration ${\cal L}$
corresponds to a (not necessarily orthonormal) basis element.

A compelling reason
to study quantum loop models is that there is already a
well-understood topological quantum field theory with loop degrees of
freedom: 
Chern-Simons theory \cite{Witten}. Here the loops are the Wilson
loops, and it is well understood how to compute their
correlators. The braiding of the strands of loop has all the 
properties for non-abelian anyons \cite{FNSWW}. In particular, if one
considers doubled Chern-Simons theory, one can obtain a
time-reversal-invariant theory. Thus one can hope that a quantum loop
model may provide a lattice version of this topological quantum field
theory. 

The connection between $SU(2)$ Chern-Simons theory and the Jones
polynomial \cite{Jones,Witten} suggests not only that loops be the
degrees of freedom, but the precise ground-state wavefunction itself
\cite{Freedman01}.  The Jones polynomial is an invariant of knots and
links. If two links have different polynomials, then they must be
topologically distinct (the converse is not true). To compute the
Jones polynomial $J(q)$, one first projects the link down on to a
plane, so that braidings become over- or under-crossings.  Each
crossing then is resolved into a sum over configurations where the
lines no longer cross. (Such a sum is particularly natural in the
context of a quantum field theory, where the space of projected links
is a Hilbert space.) Each configuration then has a value proportional
to $d^{n_{\cal L}}$, where $d=q+q^{-1}$, and $n_{\cal L}$ is the
number of loops in the collection of closed loops ${\cal L}$.

These results suggest that one study a quantum loop model where the
Hilbert space is spanned by loop configurations ${\cal L}$, and whose 
(unnormalized) ground state is
\begin{equation}
|\Psi\rangle = \sum_{\cal L} d^{n^{}_{\cal L}} |{\cal L}\rangle\ .
\label{gsw}
\end{equation}
This paper focuses entirely on models with this ground
state, even when the degrees of freedom of particular lattice realizations of
this are rewritten in terms of nets. In section \ref{sec:selfdual} I
will provide a Hamiltonian which has this $|\Psi\rangle$ as its ground
state. 

To have a finite number of ground states on the
torus, the parameter $d$ must take on special values \cite{Freedman01}.
The theory is unitary as well when $k$ is a positive integer, where
$k$ is defined by
\begin{equation}
d=2\cos\left(\frac{\pi}{k+2}\right)\ .
\label{ddef}
\end{equation}
In knot-theory language, $k$ rational corresponds to the values where
there exists a Jones-Wenzl projector \cite{Jones,JW}, which in the
quantum language is a local operator which annihilates the ground
state at the corresponding value of $d$.

This section focuses on the ``completely packed'' loop model,
where every link of the lattice is covered by self- and
mutually-avoiding loops. Here also I require that there be four links
touching each site, so two key examples are the square and Kagom\'e
lattices.  It is fairly obvious how to extend these results by
relaxing the complete-packing restraint; this  generalization
will be discussed in section
\ref{sec:generalizations}.

\begin{figure}[t]
\begin{center} 
\includegraphics[height=1.4cm]{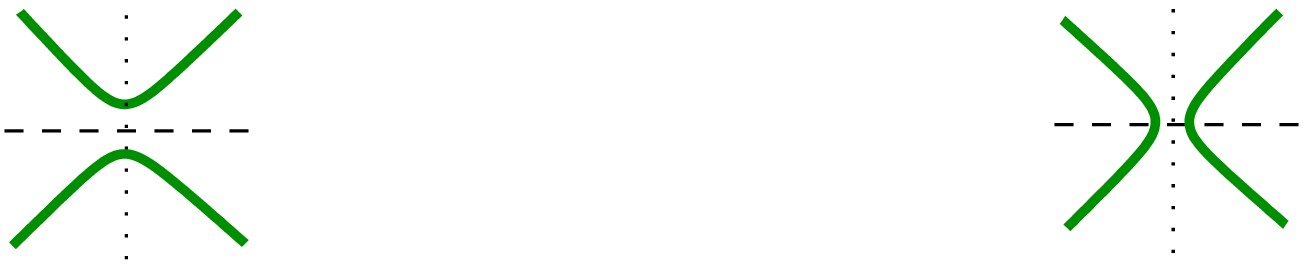}
{\large
    \put(-253,18){$|1\rangle=$}
    \put(-90,18){$\xxr=$}
}
\caption{The two-state quantum system at each vertex of the completely
  packed quantum loop model. 
The dashed line is the
corresponding link on the net lattice ${\cal N}$, while the dotted
  line is that of the dual $\Nhat$.}
\label{fig:zerodef}
\end{center}
\end{figure}

In this completely packed model, 
there are two ways that the loops can avoid each other at each
vertex. The quantum model therefore is described by a two-state system
at every vertex \cite{Freedman01,FNS}, with basis elements $|1\rangle$
and $\xxr$ illustrated in fig.\ \ref{fig:zerodef}.
Any completely packed loop configuration can be built up by choosing
one of these two basis elements at each vertex.

The {\em loop lattice} is the lattice the loops live on, where all
vertices have four links touching them.  To define which state is
which on a given loop lattice, it is convenient to define other
lattices ${\cal N}$ and $\widehat{\cal N}$.  For reasons which will
later become obvious, ${\cal N}$ is called the {\em net lattice}.  The
net lattice ${\cal N}$ is defined so that the loop lattice is the
medial lattice for both ${\cal N}$ and its dual $\widehat{\cal N}$ (a
lattice and its dual have the same medial lattice). 
In an attempt to avoid confusion with the graphs describing nets later
on, I will typically refer to the vertices of these lattices as
``sites'', and the edges of these lattices as ``links''. 
With the choice of ${\cal N}$,
$|1\rangle$ is defined so that the loops do not cross the link of
${\cal N}$; $\xxr$ is defined so that the loops do not cross
$\widehat{\cal N}$. 
When the loop
lattice is the square lattice, both ${\cal N}$ and $\widehat{\cal N}$
are also square lattices with links $\sqrt{2}$ times longer than the
links of the loop lattice. If the loop lattice is the Kagom\'e
lattice, ${\cal N}$ and $\widehat{\cal N}$ are respectively the
honeycomb and triangular lattices.

In order to define the Hilbert space of the quantum model, the inner
product must be specified. The simplest inner product is to make each
loop configuration an orthonormal basis element
\cite{Freedman01}. This amounts simply to requiring $\xxl | 1\rangle =0$.
This inner product, however, is undesirable for several reasons, and
so I refer to it as ``naive''.  

The first reason why the naive inner product is undesirable is the
``$d=\sqrt{2}$'' barrier. 
Using the naive inner product in the
completely packed loop model gives the
and the expectation value of a diagonal operator ${\cal O}$ in
the ground state is 
\begin{equation}
\frac{\langle\Psi|{\cal O} |\Psi\rangle}{\langle\Psi|\Psi\rangle}
=\frac{\sum_{\cal L} {\cal O}({\cal L}) d^{2n^{}_{\cal L}} }
{\sum_{\cal L} d^{2n^{}_{\cal L}} }
\end{equation}
where ${\cal O}({\cal L})$ is the value of ${\cal O}$ for the loop
configuration ${\cal L}$. This correlator is exactly that of a {\em
classical} loop model where each loop gets a weight $d^2$. The reason
each loop is weighted by $d^2$ and not $d$ is of course that
probabilities in quantum mechanics are given by the wavefunction's
magnitude squared.  The normalization
\begin{equation}
 \langle \Psi| \Psi\rangle= \sum_{\cal L} d^{2n^{}_{\cal L}}
\label{Zloop}
\end{equation}
is simply the partition function of this classical loop
model.

The completely packed classical loop model has long been studied. It
is the $Q=d^4$-state Potts model at its self-dual point: the loops
surround clusters in the Fortuin-Kasteleyn expansion \cite{FK}. The
loops live on the links of the medial lattice for the Potts model
lattice, so the square-lattice loop model describes the square-lattice
Potts model, while the Kagom\'e loop model corresponds to the
honeycomb or triangular lattice (since the Potts model is at its
self-dual point, the two are equivalent).  The Potts model at its
self-dual point is well understood, and the properties of the loops
can be analyzed by Coulomb-gas methods \cite{Nienhuis}. In fact, the
Temperley-Lieb algebra, of central importance in understanding
non-abelian braiding and the Jones polynomial, originally was invented
to map the square-lattice Potts model to the six-vertex model
\cite{TL}. An immediate consequence of the map is that the self-dual
point is critical only when $Q\le 4$. In terms of loops, a length
scale appears for $Q>4$ because a large-enough weight per loop favors
the creation of many short loops, instead of an ensemble of those of
all lengths. For $Q\le 4$, these completely packed loops are indeed
critical. They behave as (i.e.\ are in the same universality class as)
the loops in the $O(N)$ loop model in its dense phase, where $Q=N^2$
\cite{Nienhuis}. The behavior is the same for honeycomb and triangular
lattices as well.

These results for the classical model mean that in the quantum model,
the classical loop partition function $\langle
\Psi|\Psi\rangle$ is dominated by ``short loops'' when the weight per
loop $d^2$ is larger than $2$ \cite{Nienhuis}. Thus loops of
arbitrarily-long length do not appear in the ground state of the
quantum model.  Such behavior is not topological, since a length scale
appears. This means that anyons are confined when $d>\sqrt{2}$: there
is a ``$d=\sqrt{2}$ barrier'' \cite{Freedman01,FNS}.  Even if one
relaxes the requirement of complete packing by allowing links not to
have loops on them, the $d=\sqrt{2}$ barrier remains: the ``dilute''
$O(N)$ model does not have a critical point with all lengths of loops
when $N>2$ \cite{Nienhuis}.

Using the naive inner product is therefore undesirable because of this
barrier. Since $k=1$ has abelian statistics \cite{Kitaev97}, these
leaves only one candidate quantum loop model with the naive inner
product, that with $k=2$ and $d=\sqrt{2}$. The loops are indeed
critical, but with the naive inner product the quantum model is
gapless. A theorem \cite{FNS} indicates that there cannot be a gap
under a broad set of situations.  Moreover, it has recently been shown
that some correlators of {\em local} observables decay algebraically
as well \cite{NSTT}. A theorem \cite{Hastings} then requires that such
a model be gapless, and so quantum critical.  Such gapless models
may be quite interesting in the context of deconfined quantum critical
points \cite{Senthil}, but having a gap is necessary for obtaining
topological order.

\begin{figure}[h] 
\begin{center} 
\vskip-.02in
\includegraphics[width= .4\textwidth]{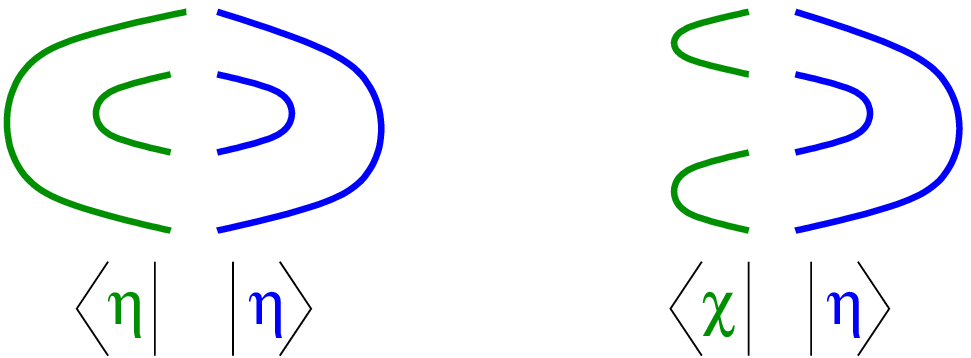} 
\caption{Gluing to find the topological inner product} 
\label{fig:gluing} 
\end{center} 
\vskip-.1in
\end{figure}

A second compelling reason why the naive inner product is undesirable
comes from considering a more general model where the loops may have
ends \cite{Levin}. If one finds a Hamiltonian so that configurations
with loop ends have a gap, then as discussed in the introduction, each
loop end is a candidate for an non-abelian anyon. Consider two such
states with loop ends at the same locations. To obtain a topological
theory like Chern-Simons in three-dimensional spacetime, one expects
that the inner product of these two states should depend only on
topological quantities. The most natural way of defining the inner
product is then to ``glue'' the dangling loop ends of the two
together, so that the combined configuration consists entirely of
closed loops. For example, let $|\eta\rangle$ and $|\chi\rangle$ each
have four loop ends in the same places, but let the loop ends be
connected in different ways.  Computing the inner product is easily
done via the schematic pictures in figure \ref{fig:gluing}.
The naive inner product requires $\langle \chi|\eta\rangle$ to vanish.
However, since  $\langle \chi|\eta\rangle$ corresponds topologically to a
single loop, it should
not vanish -- topologically it is the same as the inner product of a
two-anyon configuration with itself. This generalized model thus gives
another compelling reason to not require loop configurations to be
orthonormal. In the next sections I describe a different inner product
which solves the problems discussed in this section.

\section{Cracking the $d=\sqrt{2}$ barrier}
\label{sec:cracking}

The discussion in the preceding section illustrates that different
loop configurations should not be orthogonal. However, one cannot just
impose a topological inner product on the loops like that illustrated
in figure \ref{fig:gluing} by fiat. This would violate locality:
knowing whether two points are connected by a strand is inherently a
non-local property. Nevertheless, there is a local inner product which
allows one to go beyond $d=\sqrt{2}$, while having desirable
topological properties.

The simplest generalization of the naive inner product is to keep each
loop configuration a basis element of the Hilbert space, but not
require them to be orthogonal. In the completely packed loop model,
this means that at each vertex
\begin{equation}
\begin{pmatrix}
\langle 1|1\rangle & \langle 1\xxr \\
\xxl |1\rangle & 
 \xxl\xxr
\end{pmatrix}
=\begin{pmatrix}
1&\lambda\\
\lambda & 1
\end{pmatrix}
\label{innerprod}
\end{equation}
so that the naive inner product corresponds to
$\lambda=0$.   This inner product is positive definite
when $|\lambda|<1$.
This inner
product can be generalized to the dilute loop model in an obvious
fashion.

Keeping 
the ground state a sum over
loops with weight $d$ each  (\ref{gsw}), 
the corresponding classical loop partition function is no
longer given by (\ref{Zloop}).  With
the more general inner product, the classical partition function
$\langle \Psi | \Psi \rangle$ 
is now given by 
a sum over two completely packed loop configurations ${\cal L}$ and
${\cal L}'$ on the same lattice, one coming from $|\Psi\rangle$ and
the other from $\langle \Psi|$. Letting  $n_X^{}$ be the number of vertices
at which ${\cal L}$ and ${\cal L}'$ differ, 
\begin{equation}
\langle \Psi|\Psi\rangle 
=\sum_{{\cal L},{\cal L}'} d^{n^{}_{\cal L} +
 n_{{\cal L}'}} \lambda^{n_X^{}}\ .
\label{Zloop2}
\end{equation}

This classical partition function (\ref{Zloop2}) arising from the new inner
product (\ref{innerprod}) describes two completely packed loop models
coupled when $\lambda\ne \pm 1$ \cite{FJ}. Since the partition
function of the $Q$-state Potts model can be expanded in terms of
completely packed loops with weight $\sqrt{Q}$ each, (\ref{Zloop2})
equivalently describes two coupled self-dual $d^2$-state Potts
models. For $\lambda=1$ and $\lambda=-1$, the decoupled Potts models
are at their self-dual ferromagnetic and antiferromagnetic points
respectively, which are critical when $d=\sqrt{Q}\le 2$. However, the
inner product is not positive definite when $|\lambda|=1$, so we
cannot choose these values directly. At $\lambda=0$, the terms in the
sum in (\ref{Zloop2}) are non-zero only when ${\cal L}={\cal L}'$, so
we recover the naive inner product, where loops are weighted by
$d^2$. This is critical only when $d\le \sqrt{2}$ as described
above. At $d=\sqrt{2}$, the coupled loop models are those of the
Ashkin-Teller models, where in the spin formulation the model remains
critical for all $0\le \lambda \le 1$.

For anyons to be deconfined in the completely packed quantum loop
model, the corresponding classical loop model with partition function
(\ref{Zloop2}) must be critical for value or values of $|\lambda|<1$,
with loops of all length scales appearing in the partition
function. This phase diagram on the square lattice has been
analyzed in detail in \cite{FJ}, with the fortunate result that there
exists a critical phase even when $d>\sqrt{2}$\ ! 
It was shown
there that is a critical point at $\lambda=\lambda_c$, where
\begin{equation}
\lambda_c = -\sqrt{2}\sin\left(\frac{\pi(k-2)}{4(k+2)}\right)\ .
\label{lamc}
\end{equation}
Moreover, for $-1\le \lambda < \lambda_c$, there is a phase where the
loops are critical. In this critical phase, {\em non-local}
correlators of loops decay algebraically. An example of such a
correlator is the probability that two segments of loop are on the
same loop; this is non-local because one needs to scan the entire
system to find if there is a loop connecting the two. In \cite{FJ},
this exponent and its generalization to ``watermelon'' operators were
computed numerically, providing convincing evidence of the critical
phase. Note that this critical phase occurs for {\em negative} values
of $\lambda$, but as long as $|\lambda|<1$, the inner product
(\ref{innerprod}) in the quantum model remains positive definite.

For $k$ integer, the coupled Potts models have a local formulation in
terms of heights taking values $1,2,\dots k+2$. This arises by using
the ``RSOS'' representation of the Temperley-Lieb algebra \cite{Pasquier}. At
$\lambda=\lambda_c$, the model is critical in both the height and loop
formulations. Changing $\lambda$ away from $\lambda_c$ corresponds to
a relevant perturbation in both formulations. However, for the loops
(but not the heights), this perturbation causes a flow to the
decoupled antiferromagnetic critical points at $\lambda=-1$. 

Similar behavior also occurs in the
toric code. Indeed, below it is shown that the toric-code ground state
can be obtained at $d=\sqrt{2}$ in the completely packed loop model.
In the toric code, the corresponding classical model is bond
percolation. Correlators between individual bonds are trivially
non-critical, because the bonds are placed in an uncorrelated fashion
on the lattice. However, non-local properties of the loops
surrounding clusters of bonds are critical \cite{Nienhuis}.

Thus for $-1\le \lambda \le \lambda_c$, the classical loop model on
the square lattice with partition function (\ref{Zloop2}) is
critical. The local height model is not critical when $-1< \lambda <
\lambda_c$, so correlators of local operators decay exponentially. The
latter is a necessary \cite{Hastings} condition for having a gap. The ground
state of the quantum loop model on the square lattice for this region
of $\lambda$ therefore is
appropriate for anyonic excitations.

The corresponding classical models on other lattices do not seem to
have been studied, with one exception. This is for the ``Fibonacci''
case $k=3$, where $d=(1+\sqrt{5})/2$, the golden mean. In the next
section, I will show how this case is related to models discussed in
\cite{LevinWen,FF,Fidkowski}. A beautiful theorem from combinatorics,
Tutte's golden identity \cite{Tutte}, then implies that this classical
loop model is critical \cite{FF,Fidkowski}. Moreover, in this case,
the quantum model is closely connected to an exactly solvable
string-net model, which indeed has deconfined anyons \cite{LevinWen}.

These facts, along with general arguments based on universality,
provide good evidence that the classical loop model on other lattices
has a critical phase for $d\ge \sqrt{2}$. Thus it seems likely that
cracking the barrier should be possible for a variety of lattices, as
long as the appropriate inner product is chosen.

\section{Nets from loops}
\label{sec:nets}

In this section I show that with the appropriate choice of inner
product, the completely packed quantum loop model is naturally
described in terms of {\em nets}.  Rewriting these
loop models in terms of nets show that the ground state (\ref{gsw}) is
almost identical to that of models of 
\cite{FF}, and the simplest of the ``string-net'' models of
\cite{LevinWen}. It is thus quite likely that their excitations will
share similar properties as well.

A natural choice of $\lambda$ results from considering the fusion
properties of models with the non-abelian anyons, or equivalently, the
Wilson loops in Chern-Simons theory. The algebraic structure
underlying these fusion properties is discussed in section
\ref{sec:qga}. This choice of $\lambda$, however, can be
motivated heuristically by extending the discussion at the end of
section \ref{sec:naive}. Consider a generalized model, where loops may
have ends, on a loop lattice consisting entirely of one vertex and
four links. The two states on this lattice are simply $|1\rangle$ and
$\xxr$ at the one vertex. These two configurations are topologically
equivalent to the four-anyon configurations $|\eta\rangle$ and
$|\chi\rangle$ in figure (\ref{fig:gluing}).  When gluing these
together to obtain the inner products, it is natural to expect that
the result will be the same as that of the expectation value of the
corresponding Wilson loops in Chern-Simons gauge theories in 2+1
dimensional spacetime. Two loops are formed in $\langle\eta
|\eta\rangle$ or $\langle\chi|\chi\rangle $, while a single loop is
formed in $\langle\eta |\chi \rangle$, so the ratio of the two should
be $d^2/d= d$. Utilizing the topological correspondence with with
$|1\rangle$ and $\xxr$ means that one expects
$$
\lambda =\frac{\xxl |1\rangle}{\sqrt{\langle 1|1\rangle\,\xxl\xxr}} =
\frac{\langle\chi|\eta\rangle}{\sqrt{\langle \chi|\chi\rangle\,
    \langle \eta|\eta\rangle }} = \pm \frac{1}{d} \ .
$$
In section \ref{sec:cracking} I explained how taking $\lambda$
negative is necessary to crack the $d=\sqrt{2}$ barrier. This suggests
that the inner product (\ref{innerprod}) be defined by taking
\begin{equation}
\lambda=-\frac{1}{d}\ .
\label{lamd}
\end{equation}

This choice of inner product has remarkable consequences, resulting in
a beautiful topological and algebraic structure when the completely
packed loop model is rewritten in terms of nets. Even more
importantly, deconfined anyons are possible for $d$ larger than
$\sqrt{2}$. As shown in \cite{FJ} and discussed here in section
\ref{sec:cracking}, the corresponding classical loop model is critical
on the square lattice whenever $-1\le \lambda\le \lambda_c$, with
$\lambda_c$ given in (\ref{lamc}). This with $\lambda=-1/d$, the
loops are critical for $k\le 6$, i.e.\ $d\le 2\cos(\pi/8)$.  For
$k<6$, $\lambda=-1/d$ falls in the region where correlators of local
quantities decay exponentially. At $k=6$, $\lambda_c=1/d$, so the
local degrees of freedom are critical as well; this should therefore
describe a deconfined quantum critical point.

\subsection{The two orthonormal bases}
\label{subsec:bases}

The first step in finding nets
is rewrite the loop model in an orthonormal basis. 
The states $|0\rangle$ and $\wwr$ orthogonal to $|1\rangle$
and $\xxr$ respectively are
\begin{eqnarray}
\label{def1}
|0\rangle &=& \frac{1}{\sqrt{d^2-1}}\left(d\xxr+|1\rangle\right)\ ,\\
\wwr &=& 
\frac{1}{\sqrt{d^2-1}}\left(d|1\rangle+\xxr\right).
\label{def1tilde}
\end{eqnarray}
This indeed yields
$\langle 0|1\rangle =\wwl\xxr=0$ and $\langle 1|1\rangle
=\xxl\xxr=1.$ 
These degrees of freedom live on the sites of the loop lattice.

There are therefore two natural orthonormal bases for the Hilbert space
at each site, one with basis elements $(|0\rangle,|1\rangle)$, 
and the other with basis elements $(\wwr,\xxr)$. 
The unitary
transformation relating the vector spaces spanned by these basis
elements is implemented by the matrix
\begin{equation}
 F=
\begin{pmatrix}
\wwl|0\rangle & \wwl|1\rangle \\
\xxl|0\rangle & \xxl|1\rangle
\end{pmatrix}
=\frac{1}{d}
\begin{pmatrix}
1 & \sqrt{d^2-1}\\
\sqrt{d^2-1} & -1
\end{pmatrix}\ .
\label{Fmat}
\end{equation}
This particular matrix $F$ 
is familiar in studies of
non-abelian anyons \cite{DFNSS,Preskill}. It arises, for example, when describing the fusion of the
anyons in the Read-Rezayi series of fractional quantum Hall states
\cite{RR}. In terms of the corresponding conformal field
theory, this is the fusion matrix for two spin-1/2 primary fields in
$SU(2)_k$. 

\begin{figure}[h] 
\begin{center} 
\vskip-.02in
\includegraphics[width= .65\textwidth]{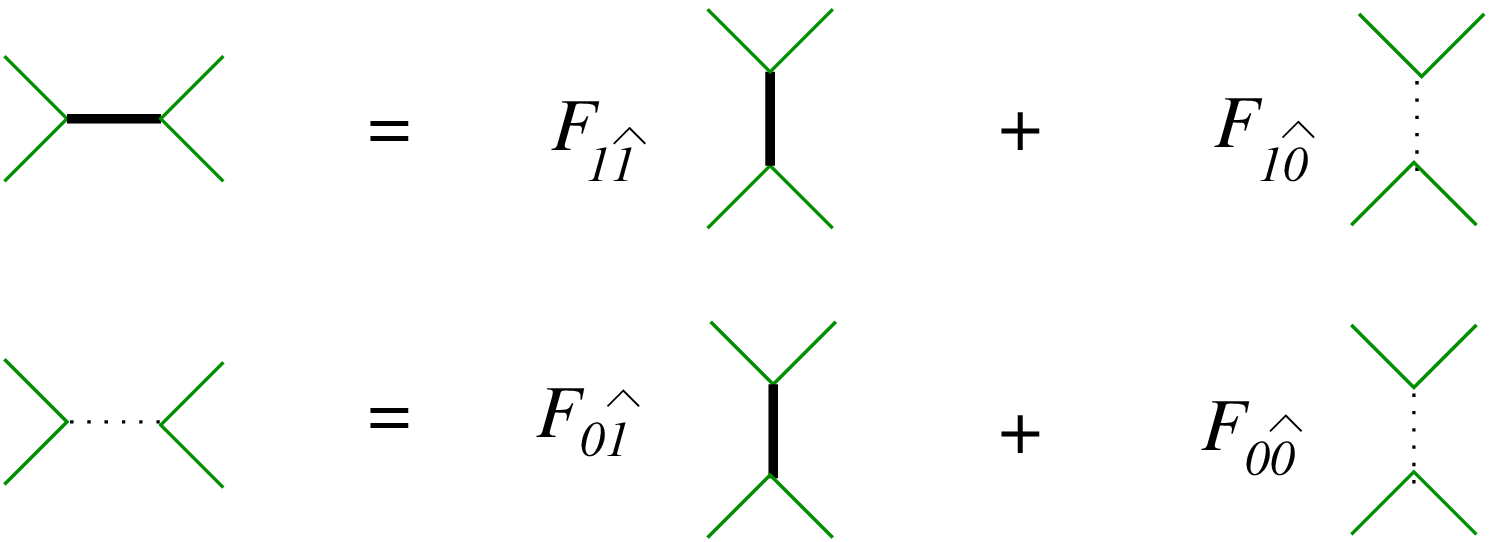} 
\caption{The four-anyon fusion matrix}
\label{fig:Fmatrix} 
\end{center} 
\vskip-.1in
\end{figure}

The correspondence of the change-of-basis matrix with the fusion
matrix is not a coincidence: at the beginning of this section it is
noted how a single vertex can be thought of as a four-anyon state in a
generalized model. The fusion matrix for non-abelian anyons is also a
change-of-basis matrix, relating different bases for four-anyon
states.  Consider four anyons at fixed positions in the identity
channel. Being in the identity channel means that the combined state
of all four is bosonic: braiding all four around anything will not
change the wave function. An essential requirement for non-abelian
statistics is that the multi-anyon Hilbert space be degenerate, so the
Hilbert space is spanned by orthonormal basis states labeled by the
index $j$. One way of specifying the basis is to pick two of the
anyons and let $j$ label the possible channels of these two anyons,
i.e.\ $j$ labels how the combination or ``fusion'' of these two anyons
will behave under braiding. Since the overall channel is the identity,
the other two anyons will necessarily be in the channel conjugate to
$j$. Labeling the four anyons by $k$, $l$, $m$ and $n$ going clockwise
from the bottom left, the basis found by fusing $k$ and $l$ is
schematically represented on the left-hand side of figure
\ref{fig:Fmatrix}.  Of course, one can choose two different anyons to
define the basis; the basis found by fusing $k$ and $n$ is shown on
the right-hand side of figure \ref{fig:Fmatrix}.
Any choice will
result in an acceptable basis, and since they all describe the same
four-anyon Hilbert space, there must be a unitary transformation
relating any two of them. The ``fusion matrix'' $F^{(klmn)}_{ij}$ is
the matrix implementing this change. This fusion can be understood
quite elegantly in terms of quantum-group algebras, and will be
explained in more depth in section \ref{sec:qga}.

Treating the change-of-basis matrix $F$ as a four-anyon fusion matrix
provides the key to rewriting the ground state of completely packed
loops in terms of nets. One orthonormal basis for the Hilbert space of
this four-anyon state is then given by $(|0\rangle,|1\rangle)$,
whereas another is $(\wwr,\xxr)$, the same basis for each vertex in
the completely packed loop model.  The change-of-basis matrix $F$ of
the latter is then indeed the four-anyon fusion matrix. Since the
fusion matrix has a nice geometric interpretation given in figure
\ref{fig:Fmatrix}, this implies a similar geometric description for
each orthonormal basis of the loop model. 

Precisely, the geometric degrees of freedom in the
$(|0\rangle,|1\rangle)$ basis live on the links of the net lattice
${\cal N}$ defined at the beginning of section \ref{sec:naive}.  Those
in the $(\wwr,\xxr)$ basis live on the links of the dual
$\widehat{\cal N}$ of the net lattice. On ${\cal N}$, one simply
denotes the state $|1\rangle$ by putting a segment of net on the
corresponding link, while $|0\rangle$ is denoted by leaving it
empty. On $\widehat{\cal N}$, one does the same for $\xxr$ and $\wwr$
respectively.
The original loop lattice can be forgotten, because complete packing
requires all its links to be covered: all the
degrees of freedom live on the links of the net lattice (or
equivalently, its dual).

\begin{figure}[h] 
\begin{center} 
\includegraphics[width= .4\textwidth]{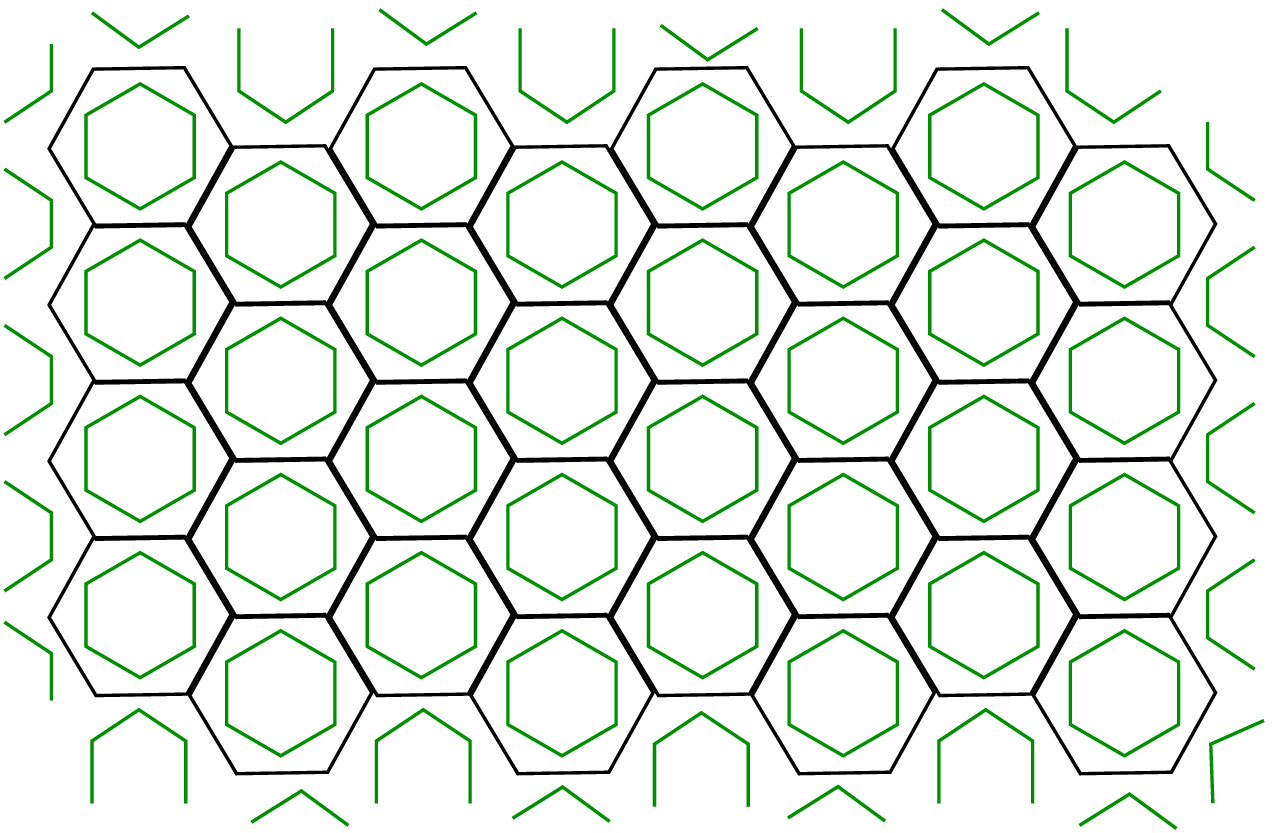} 
\caption{How the configuration with all states $|1\rangle$ is
  described as a net on ${\cal N}$}
\label{fig:kaghex} 
\end{center}
\end{figure}

\begin{figure}[h] 
\begin{center} 
\includegraphics[width= .4\textwidth]{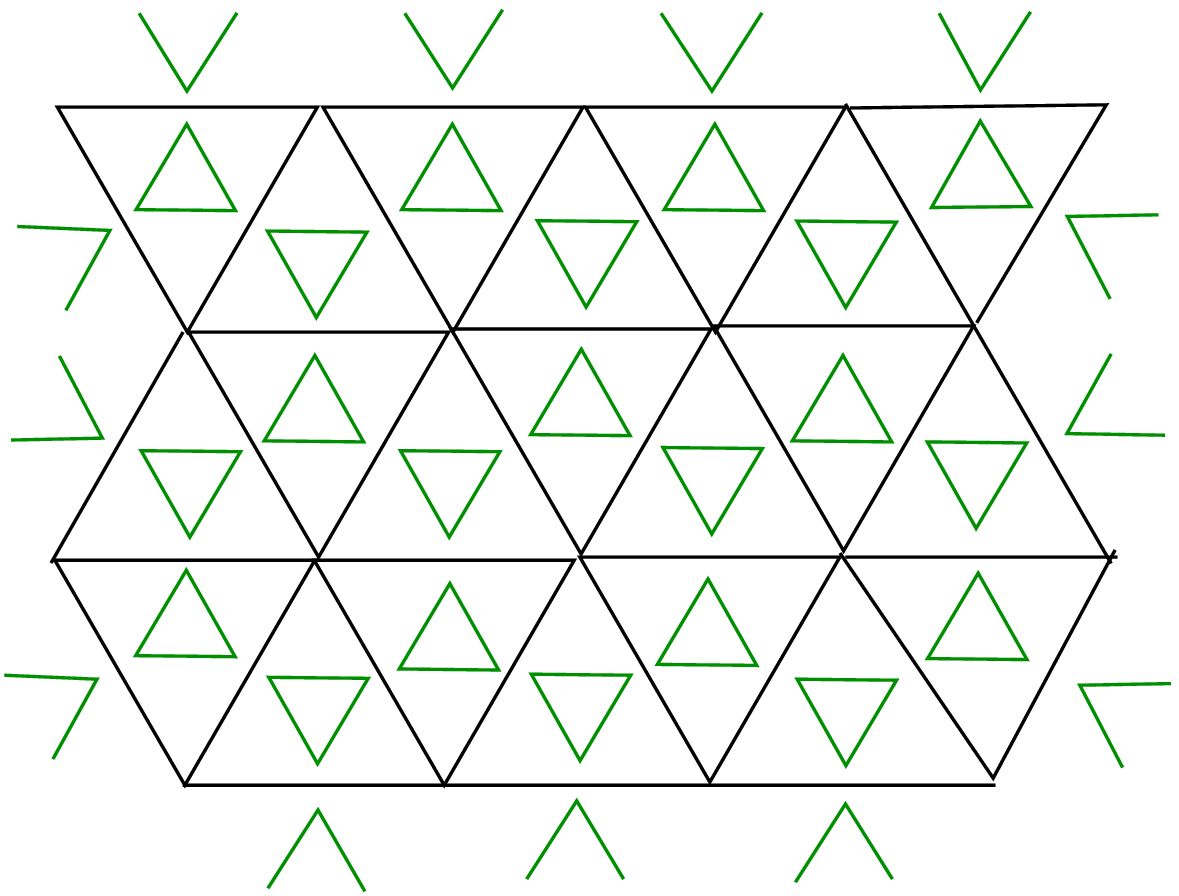} 
\caption{How the configuration with all states $\xxr$ is
  described as a net on $\Nhat$}
\label{fig:kagtri} 
\end{center}
\end{figure}

A set of orthonormal basis elements for the model can be specified by
specifying whether each link has a $|1\rangle$ or $|0\rangle$ on it;
graphically this corresponds respectively to covering the link or
not. Thus the each element $|E\rangle$ of this orthonormal basis 
corresponds to a graph $E$, whose edges are the
covered links of ${\cal N}$, and whose vertices are the sites of ${\cal
  N}$ touching these links. The example corresponding to all states
$|1\rangle$ on the sites of the Kagom\'e lattice is drawn in figure
\ref{fig:kaghex}; the graph $E$ here completely covers the net
lattice ${\cal N}$, which is honeycomb here. 
Likewise, each state written in the dual basis
$(\wwr,\xxr)$ can be represented as a graph with edges a subset of links
of $\widehat{\cal N}$. 
The example corresponding to all states
$\xxr$ on the sites of the Kagom\'e lattice is drawn in
figure \ref{fig:kagtri}, when $\Nhat$ is the triangular lattice.

\subsection{The ground state as a sum over nets}
\label{subsec:nets}

In these orthonormal bases, the ground state (\ref{gsw}) in these
bases has a beautiful topological description in terms of {\em nets}.
A net is a geometric object without ends, i.e.\ no vertex has just one
edge touching it. The beautiful part is that only nets
appear in the ground state: $\langle E|\psi\rangle = 0$ whenever $E$
has an end. 

To prove this, and to also find the topological weight of each net in
the ground state, let $|{\cal L}\rangle$ be the state of the system
written in the loop basis, i.e.\ in terms of $|1\rangle$ and
$\xxr$. Since the loop basis is not orthonormal, it is useful to
introduce the state $\langle \overline{\cal L}|$ such that for ${\cal
  L}\ne {\cal L}'$,
$$\langle\overline{\cal L}|{\cal L}'\rangle=0, \qquad
\langle\overline{\cal L}|{\cal L}\rangle=1$$
For the case at hand,
\begin{equation}
\langle\overline 1| = 
\frac{d}{\sqrt{d^2-1}}
\langle\widehat 0|, \qquad \quad
\langle\overline{\widehat 1}| = 
\frac{d}{\sqrt{d^2-1}}
\langle{0}| \ .
\label{barbasis}
\end{equation}
By construction, the weight of a particular loop configuration ${\cal
L}$ in any state $|\psi\rangle$ is then $\langle\overline{\cal
L}|\psi\rangle$. In particular, the weight in the ground state
$|\Psi\rangle$ from (\ref{gsw}) is
\begin{equation}
\langle\overline{\cal L}|\Psi\rangle=d^{n_{\cal L}} \ .
\end{equation}
Inserting a complete set of states can now be done in the loop basis:
$$\sum_{\cal L}|{\cal L}\rangle\langle\overline{\cal L}|=1\ .$$
The weight of an orthonormal basis element
in the ground state $|\Psi\rangle$ is then given by
\begin{eqnarray}
\nonumber
\langle E|\Psi\rangle &=& \sum_{\cal L} \langle E|{\cal L}\rangle
\langle \overline{\cal L}|\Psi\rangle\\ &=&
\sum_{\cal L} \langle E|{\cal L} \rangle d^{n_{\cal L}} 
\label{EPsi}
\end{eqnarray}

\begin{figure}[h] 
\begin{center} 
\includegraphics[width= .5\textwidth]{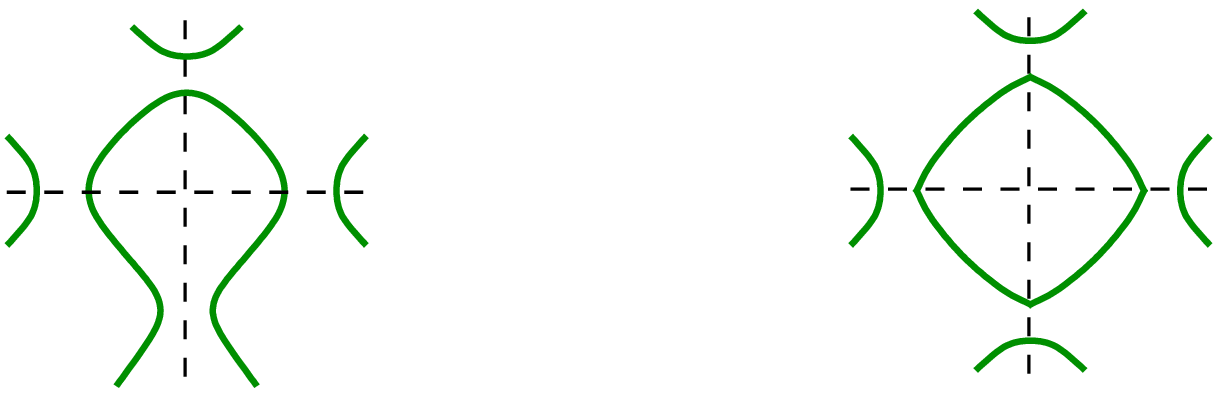} 
\caption{ The two loop configurations ${\cal L}_1$ and ${\cal L}_2$
with inner products related in (\ref{Eend}).
}
\label{fig:netend} 
\end{center}
\end{figure}

It is now straightforward to prove that if $E$ has any ends, this
amplitude $\langle E|\Psi\rangle$ vanishes. Let $E_{end}$ be any edge
configuration with an end at some vertex.  By definition, the states
on the edges around this end are of the form $\langle 100\dots|$. Thus
$\langle E_{end}|$ has non-vanishing inner product with only two
states in the loop basis:
$|1\widehat{1}\widehat{1}\dots\rangle$ and
$|\widehat{1}\widehat{1}\widehat{1}\dots\rangle$. Let $|{\cal L}_1\rangle$ and
$|{\cal L}_2\rangle$ be two loop configurations given by the former
and the latter around the vertex with the end, and otherwise
identical.  Using $\langle 1\xxr= -1/d$ means the inner products are
then related by
\begin{equation}
\langle E_{end} | {\cal L}_1\rangle = -\frac{1}{d} 
\langle E_{end} | {\cal L}_2\rangle\ .
\end{equation}
where ${\cal L}_1$ and ${\cal L}_2$ for the square lattice are
illustrated in figure \ref{fig:netend}.
${\cal L}_1$ and ${\cal L}_2$ are topologically identical,
except that the latter has an extra loop. Thus $n_{{\cal L}_2}=
n_{{\cal L}_1}+1$, and
$$\langle E_{end} | {\cal L}_1\rangle d^{n_{{\cal L}_1}} + 
\langle E_{end} | {\cal L}_2\rangle d^{n_{{\cal L}_2}}=0 \ .$$
All configurations with a non-vanishing inner
product with $\langle E_{end}|$ can be grouped in pairs of the form
of ${\cal L}_1$ and
${\cal L}_2$. Putting this information together with (\ref{EPsi})
means that
\begin{equation}
\langle E_{end} | \Psi\rangle = 0
\label{Eend}
\end{equation}
The ground state is indeed a sum over edge configurations without
ends, i.e.\ nets!

\subsection{The topological weight}
\label{subsec:weight}

The (non-vanishing) weight $\langle N|\Psi\rangle$ for each net can be found
by writing the net $N$ as a sum over loops as in (\ref{EPsi}). I
will show here that this sum can be used to find a much simpler
expression for it in terms of the chromatic polynomial. Moreover, in
this formulation it is obvious how the weight depends on the
topological properties of the nets.

The graph $N$ associated with each net is defined as the
graph whose edges are those in ${\cal N}$ which have the state
$|1\rangle$ on them, such that no vertex in ${\cal N}$ has only a
single $|1\rangle$ touching it.  The vertices of $N$ are those
vertices of ${\cal N}$ with at least two edges in $N$ touching them
(i.e.\ vertices of ${\cal N}$ with only states $|0\rangle$ touching
are not included in $N$).

To find the simpler expression for the weight of each net $\langle
N|\Psi\rangle$, it is useful to rewrite $\langle N|$ in terms
of the states in (\ref{barbasis}):
\begin{eqnarray}
\langle 1| &=& \langle\overline{1}|
-\frac{1}{d} \langle\overline{\widehat{1}}|
\label{bar1}
\\
\langle 0| &=& \frac{\sqrt{d^2-1}}{d}\langle\overline{\widehat{1}}| \ .
\label{bar2}
\end{eqnarray}
Because $\langle\overline{1}|$ does not appear on the right-hand side
of (\ref{bar2}), $\langle N | {\cal L}\rangle$ is non-vanishing only 
for loops with $\xxr$
on every edge in ${\cal N}$ not part of the net graph
$N$. 

For edges part of $N$, (\ref{bar1}) shows that both kids of loop
states contribute to the weight. Moreover, they contribute in exactly
the same combination which arose in \cite{FR,FKrush} in studies of the
Potts model and the chromatic polynomial (see e.g.\ figure 6 in
\cite{FR} or figure 8 in \cite{FKrush}). In particular, in
\cite{FKrush,FKrush2} it is shown that the chromatic
polynomial of the dual graph $\widehat{N}$ can be computed
by summing over the loop configurations created by the substitution of
the combination (\ref{bar1}) !

The precise expression for the weight of each net in the ground state
can be derived using a slight variation of Lemma 2.5 in
\cite{FKrush}. The derivation for ${\cal N}$ (and thus $N$) a planar
graph is presented in the appendix; the result is
\begin{equation}
\langle N|\Psi\rangle =
\alpha \left(\frac{1}{\sqrt{d^2-1}}\right)^{L_N}
\chi_{\widehat{N}}(d^2)
\label{netchrome}
\end{equation}
where $L_N$ is the number of edges of the net $N$, i.e.\ what in
statistical mechanics would usually be called its ``length'', $\alpha$
is an unimportant overall constant (\ref{alphad}) given in the appendix, and
 $\chi_{\widehat{N}}(d^2)$ is the
{\em chromatic polynomial} for the dual graph $\widehat{N}$.  The chromatic
polynomial vanishes for any graph $N$ which is not a net, so
(\ref{netchrome}) remains valid even if $N$ were not a net, i.e.\ one
recovers (\ref{Eend}) as well. 

The chromatic polynomial only depends on topological properties of
$N$.  Put picturesquely, if
one treats the edges of $N$ as borders separating countries,
$\chi_{\widehat{N}}(Q)$ is the number of ways of coloring each
country with $Q$ colors such that adjacent countries are not colored
the same. Typically, chromatic polynomials are described in terms of
the dual graph (the reason for the $\widehat{N}$ in the subscript of
$\chi$). The dual graph $\widehat{N}$ is defined with a vertex corresponding
to each face of $N$, and with an edge connecting two vertices for each
edge in $N$ separating the two faces. Thus when $Q$ is an
integer, $\chi_{\widehat{N}}(Q)$ counts the number of ways of
coloring $\widehat{N}$ such that any two vertices connected by an edge
have different colors. 

In statistical mechanics, the chromatic polynomial arises from the
{low}-temperature expansion of the $Q$-state Potts model. (The
completely packed loops discussed above arise from the Fortuin-Kasteleyn
expansion \cite{FK}.) The nets corresponds to domain
walls in the Potts model, separating regions of different Potts
spins. If the nets live on the links of ${\cal N}$, then these spins,
taking values $1,2\dots Q$, live on the sites of the dual lattice $
\Nhat$.  By definition, the chromatic polynomial then counts the number
of different spin configurations possible for a given domain wall
configuration. Thus the classical partition function of the Potts
model can be written as a sum over nets as
\begin{equation}
Z_{Potts} = \sum_{N} K^{L_N}\chi_{\widehat{N}}(Q) 
\label{ZPotts}
\end{equation}
where $K$ is the weight per unit length of the domain wall. In the
usual Potts language, $K\to 0$ corresponds to zero temperature and
ferromagnetic, $K\to \infty$ to zero temperature and antiferromagnetic, 
and $K=1$ to infinite temperature.

\begin{figure}[h] 
\begin{center} 
\includegraphics[width= .2\textwidth]{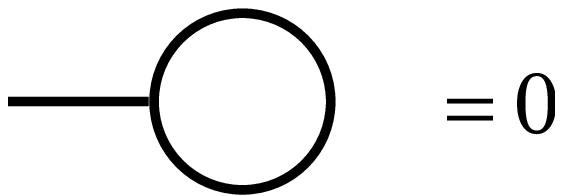} 
\caption{Nets with tadpoles do not appear in the ground state.}
\label{fig:tadpole} 
\end{center}
\end{figure}

\begin{figure}[h] 
\begin{center} 
\includegraphics[width= .4\textwidth]{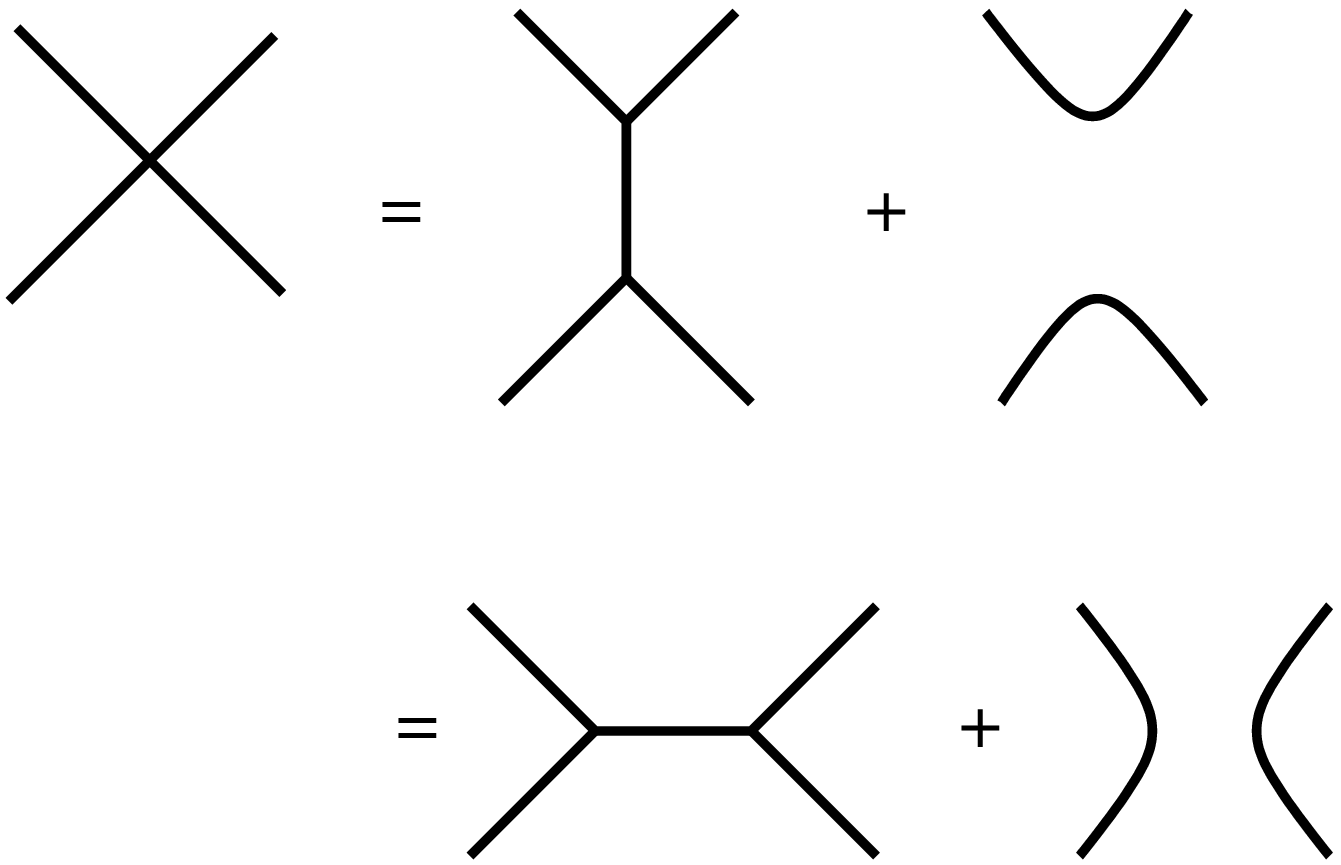} 
\caption{Relations among chromatic polynomials}
\label{fig:chromealgebra} 
\end{center}
\end{figure}

\begin{figure}[h] 
\begin{center} 
\includegraphics[width= .5\textwidth]{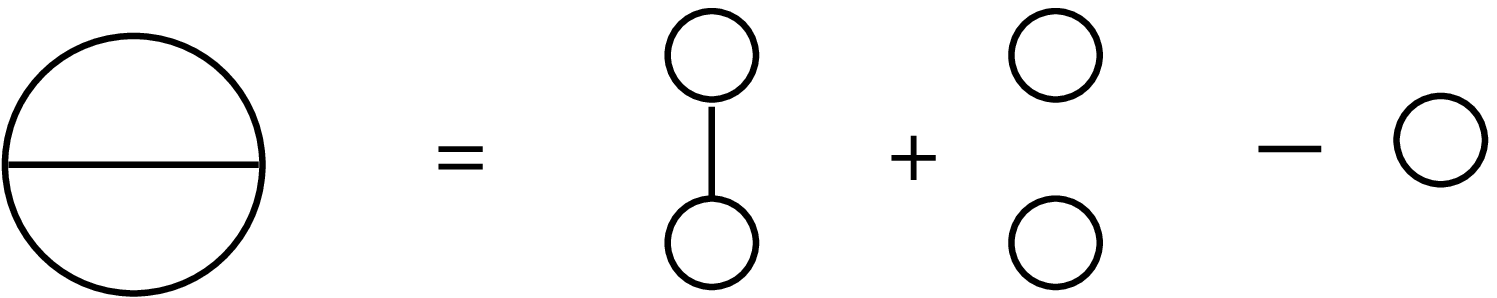} 
\caption{Evaluating the theta graph by using the relations 
in figures \ref{fig:tadpole} and \ref{fig:chromealgebra}}
\label{fig:theta} 
\end{center}
\end{figure}

It is obvious that the number of colorings for any finite graph is a
polynomial (with integer coefficients) in $Q$, so it can be
evaluated for all values of $Q$, not just integers. A common way of
computing $\chi_{\widehat{N}}(Q)$ for all $Q=d^2$ is to use
contraction-deletion relation on $\widehat{N}$: see e.g. \cite{Bolla}
or \cite{FF}. An equivalent way to define it directly in terms of $N$
for all $d^2$ is described in \cite{FKrush,FKrush2}. This method is
convenient for the study of quantum loop models, because it involves
only topological properties of $N$. One can find $\chi_{\widehat{N}}$
by repeatedly applying the following relations to simplify the graph,
and then using the fact that $\chi_{\widehat{\emptyset}}(d^2)=d^2$
for the empty graph $\emptyset$.  First, as
illustrated in figure \ref{fig:tadpole}, if any subgraph of $N$ is a
``tadpole'', $\chi_{\widehat{N}}$ must vanish. 
Second, the chromatic polynomial for a graph with a  4-valent
vertex can be reduced to two trivalent vertices by using the relation
illustrated in figure \ref{fig:chromealgebra}. 
Since there are two ways of doing this, this relation gives an
identity involving pairs of trivalent vertices as well.  Finally, any
simple closed curve in $N$ (i.e.\ any subgraph of $N$ which is
topologically a loop without intersections) can be removed if one
multiplies the resulting graph by $d^2-1$. All of these are relations
are valid for {\em any subgraph} of $N$, meaning that they can be
applied locally without regard to the rest of the graph. Proofs of
these relations, and the generalization to vertices with more legs,
are easy to find by using the coloring description.  Thus for example,
for ``theta'' graph illustrated in figure \ref{fig:theta},
$\chi_{\widehat{\theta}}(d^2)/d^2=0+(d^2-1)^2-(d^2-1)=(d^2-2)(d^2-1)$.

The upshot is that the ground state in the orthonormal net basis is
\begin{equation}
|\Psi\rangle = \alpha\sum_{N}
 \left(\frac{1}{\sqrt{d^2-1}}\right)^{L_N} \chi_{\widehat{N}}(d^2)|N\rangle
\label{Psinet}
\end{equation}
This is the ground state for the models discussed at length
in \cite{FF}, once the weight per unit length of net is fixed to be
$1/\sqrt{d^2-1}$. 

A weight per unit length not equal to 1 seems to violate the
requirement that the weight be topological.  However, as discussed in
section \ref{sec:cracking}, for $d\le 2\cos(\pi/8)$, the partition function
of the associated classical loop model $\langle \Psi | \Psi\rangle$ on
the square lattice contains contributions from loops of all sizes
\cite{FJ}. Thus (\ref{Psinet}) shows that (at least for the square
lattice) this weight is desirable. Moreover, allowing ``dilution'' for
the loops (i.e.\ relaxing the complete packing requirement)
is an {\em irrelevant} perturbation. This is characteristic of a
``dense'' phase of the classical loop model \cite{Nienhuis}, where
once the weight per unit length is large enough, the system is
dominated by configurations covering the lattice. The weight per unit
length becomes therefore becomes irrelevant in this phase. Although
the weights per unit length of the loops and nets are not the same,
they are closely related: diluting loops results in diluting nets. 
It is thus reasonable to assume that the weight
per unit length of the net in (\ref{Psinet}) is irrelevant, as long as
it is large enough (order 1).

\section{Examples of topological order}
\label{sec:examples}

\subsection{An abelian case: the toric code}

The first two non-trivial values of $k$ provide some illuminating
examples. For $k=2$ ($d=\sqrt{2}$) on the square lattice, the ground
state is quite familiar: it is that of the toric code
\cite{Kitaev97}. The weight $\chi_{\widehat{N}}(2)=2$ if $\widehat{N}$
is bipartite, and is zero if it is not. Thus no nets
with trivalent vertices can appear in the ground state, and each
vertex in ${\cal N}$ must have $0$, $2$ or $4$ adjacent edges in
$N$. Moreover, the weight per unit length for $d=\sqrt{2}$ is 1, so
each allowed net has the same weight. The ground state is the
equal-amplitude sum over all nets with no trivalent vertices. This is
indeed the ground state for the toric code. 
Configurations where the net has ends, or three edges touching a
vertex, are necessarily excitations. Although these are fractionalized
(they have mutual fermion statistics \cite{Kitaev97}), they are
abelian, a consequence of the fact of the each allowed net has the
same weight.

On general lattices, when $k=2$, the nets in the ground state must
have an even number of edges in $N$ at each vertex, and each such net
gets equal weight. When $k=2$ and the net lattice is honeycomb, the
ground state is filled with loops only. Even though the original
non-orthonormal loops on the Kagom\'e lattice have weight $\sqrt{2}$
per loop, these quantum loops on the net lattice have an orthonormal
inner product with weight one per loop. This corresponds to the
$O(1)$ loop model on the honeycomb lattice.  The phase diagram of this
model has been extensively studied \cite{Nienhuis}; for a weight per
unit length greater than $1/\sqrt{3}$, the loops are in a dense
phase. The dense phase is critical, with loops on average covering
almost all the lattice. Since the weight per unit length for the loops
on the net lattice here is 1, the orthonormal loops here are indeed
critical as necessary for topological order.

When $k=2$, the corresponding classical model is simply the Ising
model at infinite temperature on the dual lattice; the nets are the
domain walls separating regions of unlike Ising spins. This is the
special case $Q=d^2=2$ of the more general result for the
low-temperature expansion of the Potts model discussed above; just
because it is called the low-temperature expansion does not mean it is
not valid at infinite temperature. Because of the infinite
temperature, correlators of local objects have correlation length
effectively zero. However, for all lattices studied (like square and
honeycomb already mentioned), correlators of non-local objects like
nets decay algebraically. These correlators belong to the same
universality class as percolation models. It is thus natural to
conjecture that on any lattice, this model at $k=2$ has abelian
topological order. In section \ref{sec:generalizations}, I will show
that it is possible to get non-abelian topological order by allowing
vacancies into the (original) loop model.

As a quick check on this picture, consider a generalization with a
variable weight per unit length for nets in the ground state. In the
corresponding classical Ising model for the ground state, this
corresponds to varying the temperature. I argued above that varying
this coupling is irrelevant, so that the nets remain
critical. However, a quantum critical point occurs when this weight is
infinity and the net lattice is the honeycomb lattice. Here
this forbids configurations where a site has no net touching
it, and so yields the quantum dimer model, where the ``dimer'' is
on the single link touching each site without a net.
The nets indeed remain critical in the corresponding classical model,
but the local degrees of freedom (e.g.\ the dimers) also become
critical here. The quantum dimer model on the honeycomb lattice is
therefore at a quantum critical point, as is well known \cite{MSC}. In
the classical model, varying the coupling from infinity is a relevant
perturbation, but while the local degrees of freedom flow to the
trivial infinite-temperature point, the non-local loops remain
critical (but with different critical properties).  In loop model
language, this is known as the flow from the compact or fully packed
phase to the dense phase; see e.g.\ \cite{JK}. This thus provides a
nice non-trivial check that the quantum model does indeed possess
non-abelian topological order.

\subsection{A non-abelian case: Fibonacci anyons}

Non-abelian topological order occurs at the
next-simplest point $k=3$, where $d$ is the golden mean
$\tau=2\cos(\pi/5)=(1+\sqrt{5})/2$. Anyons with the fusion matrix
(\ref{Fmat}) with $d=\tau$ are known as ``Fibonacci anyons''
because the number of $n$-anyon states is the $n$th Fibonacci number
(see e.g. \cite{Preskill}). Here the model is closely related to one
of the ``string-net'' models of \cite{LevinWen}, known there as the
$N=1$ string-net model. Since this string-net model is exactly
solvable, this provides a valuable check on the results here.

In the string-net model of relevance here, there is a two-state
quantum system on every link of the honeycomb lattice, corresponding
to a segment of string net covering it or not. The ground state does
not include configurations with a site touched by only a single
segment of string net, so the ground state is a sum over nets
identical to those here. The weight of each string net in the
ground state is determined indirectly by
demanding that the weights of different configurations be related by
the appropriate fusion matrix. Here the fusion matrix is (\ref{Fmat})
with $d=\tau$, although now all the configurations involved are nets,
not the combination of loops and nets as illustrated in figure
\ref{fig:Fmatrix}.  An explicit expression for the weight of each
string net configuration in the ground state was derived in
\cite{Fidkowski}. It involves the chromatic polynomial as in \cite{FF}
and here, but instead of including weight per unit length, it has a
weight $\tau^{3/4}$ for each trivalent vertex in the net, i.e.\ the
number of sites of the net lattice ${\cal N}$ which have all three
links touching them. Denoting the ground state of the string-net model
to be $|\Psi_s\rangle$, equation (24) of \cite{Fidkowski} gives
\begin{equation}
\langle N|\Psi_s\rangle = \tau^{3V_{tri}/4}
\chi_{\widehat{N}}(\tau^2)/\tau^2
\label{stringnet}
\end{equation}
for $V_{tri}$ trivalent vertices.

A Hamiltonian can be found so that the string-net model is exactly
solvable, so it has been proved that the ground state has topological
order, and that the excited states are non-abelian anyons
\cite{LevinWen}. The close similarity of the ground state with that of
the net models developed here 
means that it is likely both have the same physics, even
with the much simpler Hamiltonian found in the next section. One strong
argument in favor of this follows from the discussion in section
\ref{sec:cracking}: the classical loop model for $k=3$ on the square
lattice is in a critical phase, and there is no evidence that the
honeycomb lattice behaves otherwise.

A second argument for topological order and deconfined anyons comes
from a remarkable property of the chromatic polynomial at $k=3$, the
{\em golden identity} \cite{Tutte,FKrush}. This identity involves the {\em square} of the chromatic polynomial at a particular argument:
\begin{equation}  
{\chi}_{\widehat N}({\tau}+2)=({\tau}+2)\, 
{\tau}^{3\,V_{tri}/2-4}\, ({\chi}_{\widehat N}(\tau^2))^2\ .
\label{golden}
\end{equation}
when $N$ involves at most trivalent vertices and closed loops.  
When computing correlators in the quantum theory, each configuration
in the orthonormal net basis is weighted by the square of
(\ref{stringnet}). Thus the golden identity gives the remarkable
simplification \cite{Fidkowski}
$$|\langle N|\Psi_s\rangle|^2 = (\tau+2) \chi_{\widehat{N}}(\tau+2)$$
so that e.g.\
$$\langle \Psi|\Psi \rangle = (\tau+2) \sum_N
\chi_{\widehat{N}}(\tau+2)\ .$$ Comparing with (\ref{ZPotts}) means
that the corresponding classical partition function is simply that of
a {\em single} $Q=(5+\sqrt{5})/2$ Potts model at infinite temperature.
Correlators in the ground state of the quantum model are those of this
single classical Potts model. Moreover, since this value of $Q$ is
less than 4, for an appropriate value of the weight per unit length
$K$, this model has a critical point. Here, $K_c= 0.51117...$ \cite{BTA}).
Thus the ground state $|\Psi_q\rangle$ is quantum critical when
\begin{equation}
\langle N|\Psi_q\rangle = \tau^{3V_{tri}/4} K_c^{L_N/2}\chi_{\widehat{N}}(\tau^2)
\label{qcp}
\end{equation}
Since the weight per unit length 1 in (\ref{stringnet}) is greater than
$\sqrt{K_c}$, the classical Potts model is indeed in its high temperature
phase, where the domain walls should be critical.

The models considered here have a ground state defined by
(\ref{netchrome}), which differs from that of the quantum-critical
ground state (\ref{qcp}) and the string-net ground state
(\ref{stringnet}) in two ways. It has weight $1$ instead of weight
$\tau^{3/4}$ per trivalent vertex, and has weight $\tau^{-1/2}$
instead of weight 1 per
unit length. Changing the weight per unit length corresponds to
changing the temperature in the classical Potts model, so this
perturbation is relevant at the critical point. Since
$\tau^{-1/2}>\sqrt{K_c}$, this perturbation here pushes the model
toward the high-temperature phase, where local degrees of freedom have
exponentially-decaying correlations, but the loops and nets are
critical.  There are no exact results for the consequence of changing
the weight per trivalent vertex. However, the honeycomb lattice is
bipartite, and if one instead includes a weight for a trivalent vertex
whenever it appears on half of the sites while leaving the weight on
the other half unchanged, it is known that there exists a critical
self-dual line \cite{Wu}.  Thus the corresponding operator in a
field-theory description is irrelevant at the critical point (there
are no marginal perturbations in the Potts models for $Q<4$). The
perturbing operator for the changing the weight on the other half of
the vertices should have the same dimension, so this operator should
be irrelevant as well.  Thus a weight per trivalent vertex is
irrelevant in the classical model.

The golden identity coupled with these results from classical Potts
models therefore gives a second argument implying that in the quantum
model when $k=3$ and the loop lattice is Kagom\'e, the ground state
$|\Psi\rangle$ from (\ref{netchrome}) has topological order. Since the
topological order is the same as that of the string-net ground state
$|\Psi_q\rangle$, it is reasonable to conjecture that there exists a
Hamiltonian such that the excited states over $|\Psi\rangle$ are
non-abelian anyons.

To conclude the discussion of the $k=3$ case on the honeycomb lattice,
it is interesting to understand why the ground state (\ref{netchrome})
for $d=\tau$ and the string-net ground state (\ref{stringnet}) are
slightly different (even though the above arguments indicate their
physics is the same). The latter arises by demanding that the fusion
matrix for the {\em nets} be unitary. 
The fusion matrix for nets $F^{\hbox{net}}$ relates
linear combinations of weights of nets in the ground state. 
In the $k=3$ case, $F^{\hbox{net}}$
relates the weights of the configurations given
in figure \ref{fig:Fmatrix}, if all the
thin green lines are replaced with thick net lines. The reason
$F^{\hbox{net}}$ is two-by-two when $k=3$ is clear from
the quantum-group language discussed below in section
\ref{sec:qga}: the spin-2 representation when $k=3$ is reducible, so two
spin-1 representations here can fuse only to spin 0 or spin 1. (This
statement can be phrased graphically via the Jones-Wenzl projector \cite{FF,FKrush}.)

The fusion matrix for nets can be computed here by using the
expression of the weights (\ref{Psinet}) in terms of the chromatic
polynomial. Using the relation in figure (\ref{fig:chromealgebra}) 
and the Jones-Wenzl projector for the chromatic polynomial at
$k=3$ \cite{FF,FKrush} shows how the ground-state weights
depend on the topology of the nets being related. 
Since $F^{\hbox{net}}$ relates nets with different lengths,
the weight
per unit length in (\ref{Psinet}) causes 
it to depend
on the lengths $L_v$ and $L_h$ of the 
internal vertical and horizontal lines on the right- and left-hand
sides of figure \ref{fig:Fmatrix}. Since $d^2-1=\tau$ for $k=3$,
\begin{equation}
F^{\hbox{net}}(L_h,L_v)=\frac{1}{\tau}
\begin{pmatrix}
1 & \tau^{2+L_v/2}\\
\tau^{-1-L_h/2} & -\tau^{(L_h-L_v)/2}
\end{pmatrix}\ .
\label{Fnet2}
\end{equation}
This matrix is not unitary, but it does satisfy the necessary
consistency requirement
$$F^{\hbox{net}}(L_h,L_v) F^{\hbox{net}}(L_v,L_h)=1\ .$$

While demanding a unitary $F$ for nets is certainly a reasonable
requirement (and seems necessary to make the string-net Hamiltonian
constructed in \cite{LevinWen} hermitian), it does not seem to be a
necessary one. The fusion matrix (\ref{Fmat}) was constructed to
change between orthonormal bases, and so in this context must be
unitary. This fusion matrix for the nets $F^{\hbox{net}}$ need not be
unitary, because it is not a change of basis: it describes a
topological property of the ground state written in terms of nets.
Consistency in this context requires only that
$F^{\hbox{net}}(L_h,L_v)F^{\hbox{net}}(L_v,L_h)=1$. However, when one
modifies the preceding calculation to compute $F^{\hbox{string net}}$
from the string-net ground-state weights given by (\ref{stringnet}),
the factors of $L_h$ and $L_v$ do not appear, and the weight per
trivalent vertex rescales the off-diagonal terms to give
$F^{\hbox{string net}}=F$ from (\ref{Fmat}). This indeed is unitary,
and moreover makes the application of the golden identity quite
natural \cite{Fidkowski}.

The crucial property of the ground state is that in the continuum,
there be no confinement scale, and I have argued at length above that
this holds even with the weight per unit length of the nets in the
ground state. This implies that it is perfectly consistent to require
that the weights of the nets in the ground state be related by using
$F^{\hbox{net}}$ in (\ref{Fnet2}) instead of $F$.  The advantage of
relaxing the requirement of unitarity is that it is much easier to
find a Hamiltonian with $|\Psi\rangle$ as its ground state than it is
to find one with $|\Psi_s\rangle$. In the next section
\ref{sec:selfdual}, I describe this Hamiltonian.

\section{Quantum self-duality and the Hamiltonian}
\label{sec:selfdual}

In this section I show that a key advantage of making nets from loops
is that such a net model is {\em quantum self-dual}. This means that
there are two different natural ways of writing the ground state
$|\Psi\rangle$, one on the net lattice ${\cal N}$ as described in
section \ref{subsec:weight}, and the other on the dual 
$\widehat{\cal N}$ of the net lattice. The quantum self-duality makes it
easy to find a Hamiltonian having $|\Psi\rangle$ as its ground
state. On the square lattice, such a Hamiltonian only requires
four-spin interactions.

The quantum self-duality of the quantum net model described in this
paper follows almost immediately from the preceding analysis.  The
choice of which lattice is the net lattice ${\cal N}$, and which is
the dual $\widehat{\cal N}$ is arbitrary: both have the
loop lattice as their medial lattice. By definition, exchanging ${\cal
N}$ with $\Nhat$ simply corresponds to exchanging the state $|1\rangle$
with $\xxr$, and so $|0\rangle$ with $\wwr$. Thus a {\em dual-lattice
net} $D$ can be defined in the same way as the net $N$. Namely, the
edges of $D$ live on the links on the dual lattice $\Nhat$, with each
edge corresponding to the state $\xxr$.  This dual-lattice net is
called $D$ to avoid confusion with $\widehat N$, the dual of a net $N$
on ${\cal N}$.

The weight $\langle D|\Psi\rangle$ of the net $D$ in the ground state
is therefore given by a computation identical to that which led to 
the expression (\ref{netchrome}). All one needs to do is exchange the
basis ($|0\rangle$, $|1\rangle$) with $(\wwr,\xxr)$; the two 
bases have equivalent properties. Rerunning the 
derivation in the appendix of the weight in this fashion gives
\begin{equation}
\langle \widehat{D} | \Psi\rangle = \alpha'
\left(
{{d^2-1}}\right)^{-L_D/2}
\chi_{\widehat{D}}(d^2)
\label{PsiNdual}
\end{equation}
Here $L_D$ is the length of the dual net $D$, i.e.\ the number of edges
covered by the state $\xxr$. The unimportant constant $\alpha'$ is
given by replacing ${\cal N}$ with $\Nhat$ in 
expression (\ref{alphad}) for $\alpha$ in the appendix.

It is important to emphasize that the ground state $|\Psi\rangle$ in
(\ref{PsiNdual}) is the {\em same} ground state being discussed all
along. This ground state can be written as a sum over nets $N$ on
${\cal N}$, or as sum over nets $D$ on the dual lattice $\Nhat$. The
weight of each state in each of these sums is defined in the same
fashion. For this reason I
refer to ground state as being {\em quantum self-dual}.  
Note that it is essential to include the weights per unit length in 
(\ref{netchrome},\ref{PsiNdual}) to have this quantum self-duality.

This quantum self-duality implies non-trivial relations among
chromatic polynomials. For example, by using the $F$ matrix change of
basis, $\chi_{\widehat{N}}$ can be rewritten as a sum over chromatic
polynomials of edge subgraphs of the dual graph. Namely, consider a
case where $N={\cal N}$, i.e.\ every link of the net lattice ${\cal N}$ is
covered.  When an edge in $N$ is replaced with a sum over unfilled and
filled edges of $\Nhat$ by using the $F$ matrix, the entries of the
$F$ matrix (\ref{Fmat}) give an unfilled edge a factor $-\sqrt{d^2-1}$
times that of the filled one. Thus
$$\chi_{\widehat{N}}(d^2)\ \propto\ \langle N|\Psi\rangle\ \propto\
\sum_{D} (-\sqrt{d^2-1})^{-L_D} \langle D|\Psi\rangle\ \propto\
\sum_D (1-d^2)^{-L_D} \chi_{\widehat{D}}(d^2)\ .
$$
where the proportional signs can be replaced with the actual
constants with a little effort.

It is useful to check this quantum self-duality by rederiving this identity 
using the duality of partition function of the classical Potts model (the
Tutte polynomial in mathematical language). Classical Potts duality
written in the low-temperature expansion (\ref{ZPotts}) is
$$Z_{Potts} = \sum_{N} K^{L_N} \chi_{\widehat{N}}(d^2) \propto
\sum_{D} (K_D)^{L_D} \chi_{\widehat{D}}(d^2)
$$ where as before $D$ is a net graph on the dual $\Nhat$ of the net
lattice, and the weight per unit length of $D$ is related to that
of $N$ by
$$\left(\frac{1}{K_D}-1\right) \left(\frac{1}{K}-1\right) = d^2\ .$$
This duality holds for any $d^2$ and any net graph ${\cal N}$. 
The chromatic polynomial for a given net $N$ can be found by taking
the (zero-temperature antiferromagnetic) limit $K\to\infty$, so that
the sum over $N$ reduces to the graph where $N={\cal N}$. When $K\to\infty$, then $K_D=1/(1-d^2)$, so indeed
$$\chi_{\widehat{N}}(d^2) \propto \sum_{D} (1-d^2)^{-L_D}
\chi_{\widehat{D}}(d^2)\ .$$ This weight per unit length is exactly
what arises from exploiting quantum self-duality. This illustrates the
quantum self-duality is quite different from classical self-duality. (Classical
self-duality is the fact that there is a value where $K=K_D\equiv K_c$
so that the Potts model on ${\cal N}$ at $K=K_c$ is equivalent to the
Potts model at $K_D=K_c$ on $\Nhat$.)

The quantum self-duality makes it easy to find a Hamiltonian $H$ with
$|\Psi\rangle$ as its ground state. The trick is to find local
projection operators that annihilate $|\Psi\rangle$. Any such operator
can be included in $H$; since by construction $H|\Psi\rangle=0$, and
eigenvalues of projection operators are non-negative, $|\Psi\rangle$
is then a ground state.  Such Hamiltonians in this context are usually
referred to as being of Rokhsar-Kivelson type \cite{RK}.  It is
essential to include enough terms in $H$ so that when space is
topologically a sphere or annulus, $|\Psi\rangle$ is the only ground
state. This should be the case if the space of states is connected by
$H$, meaning that by acting with $H$ enough, one can reach any state
in the Hilbert space from any other.

The marvelous consequence of quantum self-duality is that it gives
enough projectors to make $H$ connected. This is because any
state included in the ground state can be geometrically described as a
net on both the net lattice ${\cal N}$ {\em and} on the dual lattice
$\Nhat$. Thus any projection operator which annihilates all nets $N$
on ${\cal N}$ {\em or} annihilates all nets $D$ on $\Nhat$ also
  annihilates $|\Psi\rangle$.

Define
the projection operators $P_0$ and $P_1$ acting on each link in the
$(|0\rangle,|1\rangle)$ basis as
\begin{equation}
P_0=
\begin{pmatrix}
1&0\\
0&0
\end{pmatrix}
,\qquad
P_1=
\begin{pmatrix}
0&0\\
0&1
\end{pmatrix},
\label{Pdef}
\end{equation}
Let $V$ be a vertex of ${\cal N}$, and $e_{V1},e_{V2}, \dots$ be the
edges touching this vertex.  
Then let
\begin{equation}
{\cal P}_{V} = P_1(e_{V1}) P_0(e_{V2}) P_0(e_{V3})\dots\
 + \hbox{ rotations}
\end{equation}
where $P_i(e)$ acts on the quantum two-state system on the edge $e$,
and the $\dots$ represent $P_0$ acting on any remaining edges around $V$.
In other words, each term in ${\cal P}_V$ has $P_1$ acting on one edge
touching $V$ and $P_0$ on the others. Therefore ${\cal P}$ annihilates
any net, since by definition a net includes no vertices with only one
$|1\rangle$ touching them. The analogous operators ${\cal
P}_{\widehat{V}}$ can be defined in the $(\wwr,\xxr)$ basis, and
because of the quantum self-duality, also must annihilate
$|\Psi\rangle$. 

In the $(|0\rangle,|1\rangle)$ basis, the operators ${\cal P}_V$ are
diagonal, so a Hamiltonian comprised solely of these is not
sufficient; they annihilate any net. The key to finding a Hamiltonian
with a unique ground state is to observe that the statement that
${\cal P}_{\widehat{V}} |\Psi\rangle =0$ is independent of
basis: ${\cal P}_{\widehat{V}}$ transformed to the
$(|0\rangle,|1\rangle)$ basis still annihilates $|\Psi\rangle$.  Thus
the operators $P_{\widehat{0}}$ and $P_{\widehat{1}}$, defined by
\begin{eqnarray*}
P_{\widehat{0}} &=& FP_0 F = 
\frac{1}{d^2}
\begin{pmatrix}
1& \sqrt{d^2-1}\\
\sqrt{d^2-1}& d^2-1
\end{pmatrix}
\\
P_{\widehat{1}} &=& FP_1 F=
\frac{1}{d^2}
\begin{pmatrix}
d^2-1 & -\sqrt{d^2-1}\\
-\sqrt{d^2-1} & 1
\end{pmatrix}\ ,
\end{eqnarray*}
project onto the states $\wwr$
and $\xxr$ respectively in the $(|0\rangle,|1\rangle)$ basis.
Each vertex of $\Nhat$ corresponds to a face of ${\cal
  N}$. Letting $e_{F1},e_{F2},\dots$ be the edges around the face $F$, define
\begin{equation}
{\cal P}_{F} = P_{\widehat{1}}(e_{F1}) 
P_{\widehat{0}}(e_{F2}) P_{\widehat{0}}(e_{F3})\dots\
 + \hbox{ rotations}\ .
\end{equation}

Each of the operators ${\cal P}_V$ and ${\cal P}_F$ annihilates
$|\Psi\rangle$. Each is a projection operator, so it has only
non-negative eigenvalues. In the $(|0\rangle,|1\rangle)$ basis, each 
${\cal P}_V$ is diagonal, but ${\cal P}_F$ is not; in fact, every entry of
the latter is non-zero. When rewritten in the 
original (non-orthonormal) loop basis, all these operators contain
off-diagonal terms. Therefore the Hamiltonian
\begin{equation}
H= \sum_V {\cal P}_V + \sum_F {\cal P}_F
\label{ham}
\end{equation}
likely has all the desired properties for topological order and
non-abelian anyons in the spectrum. Exploiting the quantum
self-duality is essential to finding this $H$.  

It is illuminating to understand how the Hamiltonian acts in the loop
basis.  The matrix elements are found via
$$\langle \overline{\cal M}|H|{\cal L}\rangle = \sum_{m,n}
\langle \overline{\cal M}|m\rangle\,\langle m|H|n\rangle\,
\langle n|{\cal L}\rangle$$ 
so that the change of basis matrix $R$ for each site of the loop
lattice is 
$$R=
\begin{pmatrix}
\langle 0|1\rangle & \langle 0\xxr \\
\langle 1|1\rangle & \langle 1\xxr 
\end{pmatrix}
=\frac{1}{d}
\begin{pmatrix}
0 & \sqrt{d^2-1}\\
d & -1
\end{pmatrix}\ .
$$
$R$ is not unitary because the loop basis is not
orthonormal. 

Rewriting ${\cal P}_V$ in this basis gives an upper-right-triangular
matrix which acts non-trivially on only two kinds of loop
configurations, one of the form
$|\widehat{1}\widehat{1}\widehat{1}...\rangle$ around the vertex, and
the other $|1\widehat{1}\widehat{1}...\rangle$. For the square
lattice, these are the two types of configurations pictured in figure
\ref{fig:netend}.  Rewriting ${\cal P}_F$ in the loop basis exchanges
$|1\rangle$ with $\xxr$, but since this acts around a face of the net
lattice, the two types of loops related are still precisely those in
figure \ref{fig:netend}.  Moreover, for both ${\cal P}_V$ and ${\cal
P}_F$, the relative coefficient of the two types of terms is always
related. Namely, since
$$R^{-1}P_1 R = 
\begin{pmatrix}
1&-\frac{1}{d}\\
0&0
\end{pmatrix}
$$
then for any ${\cal M}$,
$$
-\frac{1}{d}\sum_{{\cal L}_1}
\langle{\cal M} | H |{\cal L}_1 \rangle= 
\langle{\cal M} | H |{\cal L}_2 \rangle \ ,
$$
where the sum is over all possibilities for ${\cal L}_1$, i.e.\ all
locations of the $|1\rangle$ edge around the vertex.
Since ${\cal L}_2$ has one more loop than ${\cal L}_1$, for a state 
$|\psi\rangle$ to
be annihilated by $H$,
\begin{equation}
\langle\overline{{\cal L}_2} |\psi\rangle =  
d \langle\overline{{\cal L}_1} |\psi\rangle
\label{gsloops}
\end{equation}
for each possible ${\cal L}_1$. 
Thus indeed $|\Psi\rangle$ is annihilated by $H$. 

This rewriting of $H$ to act on loops gives more than a check that
$|\Psi\rangle$ is a ground state: it gives a proof that $|\Psi\rangle$ is the
unique ground state when space is topologically a sphere. It is
straightforward to see that {\em any} two loop configurations on the
sphere can be related by repeatedly doing ``surgery'' between
configurations of the type in figure \ref{fig:netend}
\cite{Freedman01,FNS}. (This is not true on the torus, because this
surgery will not change the number of loops around the torus.) In the
ground states, (\ref{gsloops}) must be obeyed for any two
configurations related by a single surgery. So repeating this surgery
means all relative coefficients are determined, leaving $|\Psi\rangle$
as the unique ground state on the sphere.

I do not know how to prove that this Hamiltonian has a gap, but it
seems very plausible. The first term in (\ref{ham}) gives an energy to
net ends, while the second does likewise for any configuration with a
dual-lattice net end. This does not prove the model is gapped (bear in
mind the theorem of \cite{FNS}), because a net end is not an exact
eigenstate. Nevertheless, the ground state is that of the toric code
in the $k=2$ case and almost identical to that of the string-net
models for $k=3$. Moreover, local variables in the
corresponding classical models have exponentially decaying correlators
for $k<6$. Thus it is reasonable to conjecture that this Hamiltonian
is gapped.

To understand this Hamiltonian in more depth, it is useful to consider
the special case $d=\sqrt{2}$, where $k=2$. 
As discussed in section \ref{sec:examples}, for this case, the
ground state $|\Psi\rangle$ is identical to that of the toric
code. However, the Hamiltonian is not, and the reason is illuminating.

The Hamiltonian for the toric code is conveniently written in terms of
Pauli matrices acting on the $(|0\rangle,|1\rangle)$ basis. On the
square lattice it is
$$H_{toric} = \frac{1}{2}\sum_V
(1-\sigma^z_{V1}\sigma^z_{V2}\sigma^z_{V3}\sigma^z_{V4})
+ \frac{1}{2}\sum_F(1-\sigma^x_{F1}\sigma^x_{F2}\sigma^x_{F3}\sigma^x_{F4})\ .
$$ The vertex terms ensure that the ground state is a sum over
nets. The face terms are non-diagonal, but map nets to nets, and so
ensure that the ground state is a sum over nets with equal
amplitudes. The face terms only can change the length $L_N$ of the
nets by an even number. Thus on surfaces of genus $\ge 1$, the face terms
only can change the number of nets wrapped around a cycle by an even
number, leaving two possible ground states for each cycle.

To relate this to the Hamiltonian (\ref{ham}), it is useful to recall
a fact about $|\Psi\rangle$ noted in section (\ref{sec:examples}): the net
contains no trivalent vertices when $k=2$. Thus each projector
\begin{equation}
{\cal P}_V^{JW} =
P_{0}(e_{V1}) 
P_{1}(e_{V2}) P_{1}(e_{V3})P_{1}(e_{V4})
 + \hbox{ rotations}
\label{HJW}
\end{equation}
 annihilates $|\Psi\rangle$. This is known as the Jones-Wenzl
projector for $k=2$ \cite{Jones,JW}, and was discussed at length in
this context in \cite{FKrush}. Of course, the same arguments apply to
the dual nets as well, so
\begin{equation}
{\cal P}_{F}^{JW} = P_{\widehat{0}}(e_{F1}) 
P_{\widehat{1}}(e_{F2}) P_{\widehat{1}}(e_{F3})P_{\widehat{1}}(e_{F4})\
 + \hbox{ rotations}
\end{equation}
also annihilates $|\Psi\rangle$.
Since these have non-negative eigenvalues, adding them to the
Hamiltonian preserves $|\Psi\rangle$ as a ground state. Using
$$2P_0 = 1+\sigma^z, \qquad 2P_1 = 1-\sigma^z, \qquad 2P_{\widehat{0}}
= 1+\sigma^x,\qquad 2P_{\widehat{1}}=1-\sigma^x, $$
gives
\begin{equation}
H_{toric} = 
\sum_V ({\cal P}_V + {\cal P}_V^{JW}) +
\sum_F ({\cal P}_F + {\cal P}_F^{JW})
= H_{k=2} + \sum_V  {\cal P}_V^{JW} +
\sum_F {\cal P}_F^{JW}\ .
\label{toricH}
\end{equation}
Thus $H_{toric}$ differs from $H$ simply by adding in the
Jones-Wenzl projectors. 

There exists a unique Jones-Wenzl projector at any value of
$k$ (see the appendix by Goodman and Wenzl in \cite{Freedman01}, and
for the net case, \cite{FKrush}). Thus an interesting question is if
adding Jones-Wenzl projectors allows a model
with the same ground state to be solved for other values of $k$. For
higher values of $k$, the Jones-Wenzl projector requires involving
more nets, so the Hamiltonians will require interactions with more
spins. For $k=3$, one can find such terms involving six spins on
the square lattice, but the closely-related exactly solvable
Hamiltonian of \cite{LevinWen} hints that 12-spin interactions may be
necessary to find a solvable model.

As opposed to $H$, the Hamiltonian $H_{toric}$ is exactly solvable,
and it is easy to see that $|\Psi\rangle$ is the unique ground state
on the sphere.  The proof above indicates that this remains true for
$H$, so that this proves that the Hamiltonian of the toric code can be
deformed without changing the ground state on the sphere. An
interesting open question is whether the excited states remain anyonic
and gapped after the deformation, but the arguments above indicate
that this is a good possibility. Moreover, the existence of
Jones-Wenzl projectors is a good sign that a gapped
topological phase is possible for any $k$. Since one can add a Jones-Wenzl
projector to the Hamiltonian and preserve $|\Psi\rangle$ as the ground
state, it seems highly unlikely that one can add such a term with
arbitrary coefficient without the model having a gap. In fact, this is
essentially what the string-net models do; their Hamiltonian includes
the Jones-Wenzl projector.

\section{Quantum-group algebras in quantum loop and net models}
\label{sec:qga}

The structure of the loop and net models discussed above can be
understood very naturally in terms of {\em quantum group algebras}.
This approach also illuminates why it is
desirable to choose the inner product as above.
 
To follow this, one doesn't need to know much more than a paragraph's
worth of results on quantum groups, so these results are presented
here. For an excellent in-depth discussion in a closely related
context, see \cite{Bais}. A quantum-group algebra $U_q(G)$ is a
one-parameter ($q$) deformation of a Lie algebra $G$. The algebra
$U_q(G)$ has representations whose tensor-product rules corresponding to
representations of $G$, with one major distinction: if $q$ is a root of unity,
all but a finite number of these are reducible.  The quantum loop and
net models discussed above are related to $U_q(SU(2))$, with
$q=e^{i\pi/(k+2)}$; only the spin-$j/2$ representations with
$j=0,1,\dots k+1$ are irreducible. The
Jones-Wenzl projector discussed above annihilates the ground state
precisely because of this truncation; for the appropriate value of
$k$, the spin-$(k+1)/2$ representation can be
decomposed into representations of the irreducible ones, and so there
is a linear relation among ground-state configurations.

To give a few simple but important examples, the tensor product of two
spin-1/2 representations of $U_q(SU(2))$ with $k\ge 2$ is decomposed
into irreducible representations as
\begin{equation}
\frac{1}{2} \otimes \frac{1}{2} = 0 \oplus 1\ ;
\label{spinhalf}
\end{equation}
for $k=1$, the truncation means that the $1$ does not appear.  A
spin-1 representation has a tensor product with another spin 1 like
that of ordinary $SU(2)$ when $k\ge 4$:
\begin{equation}
1 \otimes 1 = 0 \oplus 1 \oplus 2 \qquad\hbox{ for }k\ge 4;
\label{spin1}
\end{equation}
for $k=2,3$ the $2$ does not appear, and for $k=2$ the $1$ does not
appear as well. 
The Jones-Wenzl projector for $k=2$ has already been
given in (\ref{HJW}). In quantum-group language it arises because
$1\otimes 1 \otimes 1 = 1\otimes 0 = 1$ when $k=2$, so that three
spin-1 representations do not have the identity in their tensor product.

The relation of quantum nets to quantum-group algebras with is that
each {\em type} of geometric degree of freedom in the former
corresponds to a {\em representation} of the latter.  Fusion in the
former corresponds to a tensor product in the latter, with the fusion
matrix comprised of the 6$j$ symbols.  The (chiral) primary
fields of the conformal field theory and the types of Wilson loops of
(undoubled) Chern-Simons theory are in one-to-one correspondence with
these representations.

In the $SU(2)_k$ theories considered above, each segment of loop
corresponds to the spin-1/2 representation of $U_q(SU(2))$.  The
requirement that these strands form closed loops then can be
understood as requiring that at each vertex, all four segments must
fuse to the identity representation.  Label the four strands (or
equivalently, the four spin-1/2 representations) going clockwise
around a vertex in the completely packed loop model
$\alpha,\beta,\gamma,\delta$. Consider two of them, say $\alpha$ and
$\beta$, on opposite sides of the link of the net lattice going
through this vertex. Because of
(\ref{spinhalf}), these two can fuse either to the spin-0 identity
representation, or the spin-1 representation. The other two states,
$\gamma$ and $\delta$, also can fuse to spin 0 or spin 1. The only way
for all four to fuse to spin $0$ is then for $\gamma$ and $\delta$ to fuse
in the same way that $\alpha$ and $\beta$ did. There are thus two states in
the quantum loop model, corresponding to these two choices. These are
precisely the states $|0\rangle$ and $|1\rangle$ defined above, thus
explaining the choice of notation. This procedure is no different from 
finding singlets of four $SU(2)$ spins.

Instead of fusing first the pair $\alpha$ and $\beta$, one of course
could have fused $\alpha$ and $\delta$ first. The same arguments
apply, but since this is a rotation by $90$ degrees of the earlier
situation, the states must be rotated as well. Thus this way of fusing
results in the states $\wwr$ and $\xxr$.  These two states are not
independent of $|0\rangle$ and $|1\rangle$, but are merely another
basis. In the $SU(2)_k$ conformal field theory, or in level $k$
$SU(2)$ Chern-Simons field theory, the fusion matrix (\ref{Fmat}) is
the unitary matrix transforming between these two bases. Thus setting
$\lambda=-1/d$ in (\ref{innerprod}) gives precisely the inner product
that preserves the quantum-group structure.

Understanding these fusion rules explains why the $k=2$ is equivalent
to the toric code. The net degrees of freedom are spin 1, and the
fusion rule for $k=2$ is $1\otimes 1 =0$. This is abelian, because
there is only one irreducible state on the right-hand-side. Likewise,
for $k=3$, the truncation means that fusion rule is $1\otimes 1= 0 +
1$, characteristic of the fusion of Fibonacci anyons. The truncation
also explains why $F^{\hbox{net}}$ in (\ref{Fnet2}) is $2\times 2$
instead of $3\times 3$.

It is important to note that the quantum-group algebra is {\em
not} a symmetry algebra of the model: the loops are not in a spin-1/2
(i.e.\ two-dimensional) representation of $U_q(SU(2))$. Rather, they
{\em correspond} to the spin-$1/2$ representation, meaning for example
that the loops obey the same fusion rules as do fields in this
representation of the $SU(2)_k$ conformal field theory.

This picture gives an elegant way of characterizing a ground
state. The unique feature of a ground state is that {\em at any
site of the loop, net or dual net lattices}, the corresponding
degrees of freedom on the links touching the vertex must {\em must
fuse to the identity representation}. With the way the model was
built, this is automatic for sites of the loop lattice. For the net
or dual net lattice, what this means precisely is that the ground
state is annihilated by any projector onto a fused state orthogonal to
the identity. The projectors in the Hamiltonian (\ref{ham}) are of
course examples of this, projecting onto the spin-1 fused state on
vertices of the net and dual net lattices.

The types of anyonic excitations themselves also correspond to
representations of the quantum-group algebra in the same way that the
geometric degrees of freedom do. Any time a
state is not annihilated by one of the aforementioned projectors, it
by construction is part of an excited state. This
is the precise meaning of ``cutting'' the net. There will then
be anyons corresponding to all irreducible representations of the
quantum group that can be built up by taking tensor products of the
representations corresponding to the nets. Thus even though there
are half-integer-spin representations of $U_q(SU(2))$, in the model
discussed above only integer-spin anyons occur because segments of
nets themselves correspond to the spin-1 representation.

\section{Generalizations and Conclusions}
\label{sec:generalizations}

This paper introduces and develops models on any lattice whose ground
state is a sum over all net configurations. When $k<6$ and the lattice
is square, the work of \cite{FJ} indicates that correlators of
non-local objects like nets in the ground state decay algebraically,
while correlators of local objects decay exponentially. Although this
does not guarantee that the model is in a gapped non-abelian
topological phase, this is a necessary condition. The similarities
between these models and the string-net models on the honeycomb
lattice also provide substantial evidence for topological order.

One new aspect of the work here is the concept of quantum
self-duality, which made it possible to construct a simpler
Hamiltonian annihilating the ground state. The key to finding a
quantum self-dual ground state was to ensure that the weight of each
configuration on the loop lattice, even if not orthonormal, came from
the corresponding link invariant. In the case above, the link
invariant was the Jones polynomial, but one can construct link
invariants (see e.g. \cite{Kuperberg}) for any $U_q(G)$. Then
combining this ground state with the orthonormal basis constructed
using the $F$ matrix allows the weights of each orthonormal basis
element in the ground state to be found. Since by construction, the
nets will still fuse to the identity representation at each vertex,
the quantum-group structure should ensure that a nice Hamiltonian with
this ground state acting on the nets can be found.

The quantum net models developed here arise
from completely packed loops on the medial lattice. There
are a number of ways of relaxing this restriction, culminating in the
string-net models of \cite{LevinWen}, where each allowed state on a
link corresponds to a primary field in a rational conformal field
theory. These generalizations are straightforward to make by utilizing
the quantum-group algebras discussed in section \ref{sec:qga}, and
following the work of e.g.\ \cite{Kuperberg} and \cite{LevinWen}. One
specifies the loop lattice, and then which representation(s) of some
quantum group $U_q(G)$ appear on which links. Thus for example, in
\cite{LevinWen}, each link of the honeycomb lattice can be in any
representation of the quantum-group algebra. The quantum-group algebra
will then provide a fusion matrix which allows an orthonormal basis to
be determined, the guiding principle being that different
representations at the same place (site or link) must be
orthogonal. The ground state involves only configurations which fuse
to the identity at every vertex. The associated link invariants then
provide the topological part of the weight of each configuration in the
ground state.

Following this strategy, it appears possible to construct a model with
non-abelian anyons at $d=\sqrt{2}$ as well as at larger values. This
is done by allowing links of the loop lattice to be empty, i.e.\
allowing the spin-0 representation of $U_q(SU(2))$ as well as the
spin-1/2. Then there are degrees of freedom on both the links and
sites of the loop lattice. The ground state will still remain a sum
over loops with a weight $d$ per loop, as in (\ref{gsw}), only now
loops are not completely packed. The inner product at sites with all
four edges covered by loops will also remain the same. The
corresponding classical loop model will essentially the same, only now
there will be a fugacity governing the amount of dilution. Since
dilution is irrelevant in the classical coupled loop models \cite{FJ},
it seems likely that this quantum model will be in a deconfined phase
for $k<6$ like the completely packed version. The interesting feature
of this model over the completely packed version is that there are
excitations of spin $1/2$ (and $3/2$), at sites on the loop lattice
with an odd number of loops touching them. The spin-1/2 ones are
non-abelian even at $d=\sqrt{2}$, because the fusion algebra
(\ref{spinhalf}) remains non-trivial here.

One can of course generalize the models of this paper without relaxing
the complete packing constraint by still requiring that each link of
the loop lattice be in the same representation of $U_q(SU(2))$, but
using a representation with spin greater than $1/2$. For
spin 1, the fusion relation (\ref{spin1}) then requires that there be a
three-state quantum system at each site (when $k>3$). Following the
analysis of \cite{FR,FF,FKrush,FKrush2} shows that the ``loops''
here will be nets, with weights given by the chromatic polynomial.

This construction does not guarantee quantum self-duality,
nor does it guarantee that correlators in the ground state will have
the desired properties for topological order. However, the fact that
it works (at least for $k<6$) in the above models, and always works in
the string-net models, means it is very reasonable to hope that this
construction of simple(r) Hamiltonians will work in more general
settings.

\bigskip

I am grateful to Eduardo Fradkin, Jesper Jacobsen, Slava Kruskhal and
Nick Read for collaborations on papers that were essential to this
work.  I also would like to thank Michael Freedman, Roderich Moessner,
Chetan Nayak, Kirill Shtengel, and Shivaji Sondhi for many
conversations on quantum loop models. This research has been supported
by the NSF under grants DMR-0412956 and DMR/MSPA-0704666, and by an
EPSRC grant EP/F008880/1.

\appendix
\section{Deriving the topological weight}

In this appendix I adapt Lemma 2.5 of \cite{FKrush} to show how the
weight of each net in the ground state can be expressed in terms of
the chromatic polynomial. As in (\ref{EPsi}), inserting a complete set
of states gives $\langle N|\Psi\rangle$ as a sum over loops:
\begin{equation}
\langle N|\Psi\rangle =
\sum_{\cal L} \langle N|{\cal L} \rangle\, d^{n_{\cal L}} \ .
\label{NPsi}
\end{equation}
As discussed above, because of (\ref{bar2}), $\langle N|{\cal L} \rangle$
is nonzero only when the states on
all edges of ${\cal N}$ not part of the net $N$ are fixed. The sum over loops here
therefore reduces
to a sum over subgraphs $N_s$ of $N$, depending which of the terms in
(\ref{bar1}) contributes on each edge. Each $N_s$ has the same
vertices as $N$, but a subset of edges, defined as those
where the {\em second} term in (\ref{bar1}) contributes.
Each $N_s$ therefore corresponds to a single loop configuration.
Therefore
\begin{equation}
\langle N | \Psi\rangle = \sum_{N_s\subseteq N}
\left(-\frac{1}{d}\right)^{E(N_s)}
\left(\frac{\sqrt{d^2-1}}{d}\right)^{E({\cal N})-E(N)} d^{n_{{\cal L}_s}}
\label{NPsinL}
\end{equation}
 where $E(G)$ and $V(G)$ are the number of edges and vertices in
the graph $G$, and $n_{{\cal L}_s}$ is the number of loops in the unique
loop graph determined by $N_s$. 

The next task is therefore to relate the number of loops $n_{\cal L}$
to topological properties of $N_s$. Because the loop state is always
$\xxr$ on all edges not part of $N$, each vertex of ${\cal N}$ not
part of $N$ is surrounded by a single loop of minimum length.
Thus the number of loops
surrounding vertices of ${\cal N}$ not in $N$ is $V({\cal N})-V(N)$.
It is straightforward to see by drawing pictures (see 
figure 12 in \cite{FKrush}) that the
remaining loops are the boundary of a regular neighborhood of
$\widehat{N_s}$, the dual graph of $N_s$.  The total number of such
loops is $k(s)+n(s)$, where $k(s)$ is the number of connected
components of $\widehat{N_s}$, and $r(s)$ is the rank of the first
homology of $\widehat{N_s}$ (i.e.\ the number of independent
cycles). Therefore
$$n_{{\cal L}_s} = V({\cal N})-V(N)+k(s)+n(s)\ .$$ This can be
simplified by noting that
$k(s)-n(s)+E(\widehat{N_s})=V(\widehat{N})$. This is proved by first
noting that it is true for the case where $\widehat{N_s}$ has no
edges, so that $k(s)=V(\widehat{N})$ and
$n(s)=E(\widehat{N_s})=0$. Adding an edge to $\widehat{N_s}$ then
either decreases $k(s)$ by 1 (if it connects two previously
disconnected clusters), or it increases $n(s)$ by 1 (if it adds an edge
connecting two already-connected vertices, it increases the number of
independent cycles by 1). Thus the identity follows by induction,
giving
$$n_{{\cal L}_s} = V({\cal N})-V(N)+2k(s)+ E({N_s})-V(\widehat{N})\ ,$$ 
where I exploited the fact that the edges on a graph are in one-to-one
correspondence with those in the dual graph so that
$E(\widehat{N_s})=E(N_s)$.

Plugging this in to (\ref{NPsinL}) gives
$$\langle N|\Psi\rangle =
d^{r}
\left(d^2-1\right)^{(E({\cal N})-E(N))/2}
\sum_{{N_s}\subseteq {N}}
\left(-1\right)^{E(N_s)} d^{2k(s)}\ ,
$$
where
$$
r=V({\cal N})-E({\cal N})+E(N)-V(N)-V(\widehat{N})\ .
$$
Each vertex of a graph corresponds to face of the dual graph, so
$V(\widehat{N})=F(N)$, where $F(G)$ is the number of faces of $G$. 
Euler's relation $V(G)-E(G)+F(G)=2$ then gives
$$r= -F({\cal N})\ .$$ 

Finally, this can be related to the chromatic polynomial by using the
identity for the chromatic polynomial of any planar graph (see
e.g. \cite{Bolla})
$$\chi_{\widehat{N}}(d^2) = \sum_{\widehat{N_s}\subseteq \widehat{N}}
\left(-1\right)^{E(\widehat{N_s})} d^{2k(s)}\ .
$$ 
This sum over edge subgraphs $\widehat{N_s}$ of $\widehat{N}$ is
equivalent to the sum over subgraphs $N_s$ of $N$, since the edges of
a graph and its dual are in one-to-one correspondence. Thus setting
\begin{equation}
\alpha = d^{-{F}({\cal N})} (d^2-1)^{E({\cal N})/2}
\label{alphad}
\end{equation}
gives the inner product (\ref{netchrome}) for any planar net $N$. As a
quick check on this result, this computation can be done easily for
the square lattice by utilizing the Temperley-Lieb algebra. This is
done at the end of section 2 of \cite{FJ}, where the weight
$1/\sqrt{d^2-1}$ per unit length of net is indeed recovered.


\def\cmp#1#2#3{Comm.\ Math.\ Phys.\ {\bf #1}, #2 (#3)}

\end{document}